\documentstyle[11pt]{article}
\begin{document}

\newcommand{\ket}[1] {\mbox{$ \vert #1 > $}}  \newcommand{\bra}[1] {\mbox{$ <
#1 \vert $}}  \newcommand{\bk}[1] {\mbox{$ < #1 > $}}  \newcommand{\scal}[2]
{\mbox{$ < #1 \vert #2 > $}}  \newcommand{\expect}[3] {\mbox{$ \bra{#1} #2
\ket{#3} $}}  \newcommand{\ki}{\mbox{$ \ket{\psi_i} $}}
\newcommand{\bi}{\mbox{$ \bra{\psi_i} $}}  \newcommand{\p} \prime
\newcommand{\e}{\mbox{$ \epsilon $}}  \newcommand{\la} \lambda

\newcommand{\om}{\mbox{$ \omega $}}  \newcommand{\cc}{\mbox{$\cal C $}}
\newcommand{\w} {\hbox{ weak }} \newcommand{\al}{\mbox{$ \alpha $}}

\newcommand{\be}{\mbox{$ \beta $}}

\overfullrule=0pt \def\sqr#1#2{{\vcenter{\vbox{\hrule height.#2pt
          \hbox{\vrule width.#2pt height#1pt \kern#1pt
           \vrule width.#2pt}
           \hrule height.#2pt}}}}
\def\square{\mathchoice\sqr68\sqr68\sqr{4.2}6\sqr{3}6} \def\lrpartial{\mathrel
{\partial\kern-.75em\raise1.75ex\hbox{$\leftrightarrow$}}} \begin{flushright}
ULB-TH 94/02\\ April 1994\\DEFINITIVE VERSION
\end{flushright} \vskip 1.5 truecm
\centerline{\LARGE\bf{From Vacuum Fluctuations to Radiation:}}
\centerline{\LARGE\bf{Accelerated Detectors and Black Holes.}}
\vskip 1. truecm
\vskip 1. truecm
 \centerline{  S. Massar\footnote{e-mail:
smassar @ ulb.ac.be}$\/^{,}$ \footnote{Boursier IISN}} \centerline{Service de
Physique Th\'eorique, Universit\'e Libre de Bruxelles,} \centerline{Campus
Plaine, C.P. 225, Bd du Triomphe, B-1050 Brussels, Belgium} \vskip 5 truemm
\centerline{R. Parentani\footnote{e-mail: renaud @ vms.huji.ac.il}}
\centerline{
  The Racah Institute of
Physics,The Hebrew University of Jerusalem,} \centerline{Givat Ram Campus,
Jerusalem 91904, Israel}
\vskip 1.5 truecm
\vskip 1.5 truecm
\vskip 1.5 truecm
{\bf Abstract }
The vacuum fluctuations that induce the transitions and the
thermalisation of a uniformly
accelerated two level atom are studied in detail. Their energy content is
revealed through the weak measurement formalism of Aharonov et al.
It is shown that each time the detector makes a transition it radiates
a Minkowski photon.
The same analysis is then applied to the conversion of vacuum
fluctuations into real quanta in the context of black hole radiation.
Initially these fluctuations are located around the light like geodesic
that shall generate the horizon and carry zero total energy. However
upon
exiting from the star they break up into two pieces one of which
gradually
acquires positive energy and becomes a Hawking quantum, the other,
its ''partner", ends up in the singularity. As time goes by the vacuum
fluctuations
generating Hawking quanta have exponentially
large energy densities. This implies that  back reaction effects
are large.

\vfill \newpage

\section{Introduction}

Pair creation in a strong external field is a well known aspect of
quantum matter field theory. For instance, in a constant electric field,
$e^+ e^-$ pairs are spontaneously created \cite{schwing}.
 Another famous example is
the Hawking flux engendered by the time dependent geometry of an
incipient black hole \cite{hawk2}.

At present the back-reaction of these quanta on the external
field which produces them is far from being understood.
The
semi-classical treatment
alone does not give rise to difficulty.
This is because the external field remains purely classical since only
the mean
value of the matter current operator acts on it as a source
\cite{mottola} \cite{Bardeen} \cite{PT}.
All the quantum
properties of the matter, including its fluctuations, are
completely ignored by the external
field. When the fluctuations become important this fluid
description fails
and a more quantum
mechanical treatment of the back-reaction is needed.

Since a fully quantum description seems beyond the
present scope
of quantum field theory (this is certainly the case
for the gravitational back-reaction since no renormalisable theory
exists yet), an intermediate approach wherein the
quantum fluctuations are at least partially
taken into account is required.

In \cite{bmpps}, such a treatment
 based on the weak measurement formalism of Aharonov et al.\cite{aharo}
was proposed in the context of electroproduction. By
isolating through a post-selection that
part of the wave function which contains a specific pair of quanta,
the weak value of the current operator was computed and interpreted.
A clear picture of the creation act of the selected pair was obtained:
It emerges out of a vacuum fluctuation which gets progressively
distorted
by the electric field and finally converted into a pair of asymptotic
quanta. This formalism yields,
in addition to an explicit evaluation of
the fluctuations around the mean value, the first order quantum
modification of the external field. This modification in turn governs the
back reaction of the selected pair onto itself and the following ones.

The purpose of the present paper is to apply
the same construction to uniformly accelerated systems and to black hole
radiation. A detailed description of the vacuum fluctuations which give
rise to asymptotic quanta is obtained.

In order to carry out this program a generalisation of the post-selection
used in  \cite{bmpps} is required for the following
reasons. First,
in the black hole problem, incomplete post-selections are needed
since the "partners" of the Hawking quanta
are inaccessible to any asymptotic observer \cite{Wald} \cite{pbt}.
 Secondly, when a large
number of quanta are produced (in the mean), it is  unphysical to
post-select states wherein a
single pair is present
because the probability that they occur is exponentially small.
Thirdly, since post-selection is a rather formal and arbitrary
operation,
one may question the physical relevance of the resulting weak-values.

The first part of this paper (chapters 2, 3 and 4) is devoted to overcome
these
difficulties. We work in the context of post-selecting Rindler quanta
(rindlerons) in Minkowski vacuum. In particular, in chapter 4, we show how the
EPR correlations between the quantum jumps of an accelerated two-level atom
\cite{Unruh}
and the state of the radiation field coupled to it give back, in a very
natural manner, the amplitudes previously post-selected by hand.
These correlations indicate that, by
getting excited, the atom has selected out of Minkowski vacuum the
fluctuation which contains the rindleron needed to make the transition.
This vacuum fluctuation admits two complementary descriptions. An inertial
observer would say that it carries zero energy whereas a uniformly
accelerated observer in the
quadrant
of
 the atom would assign it a
positive energy.
The relation between these two interpretations is
at the heart of all our analysis.

In chapter 5, we reverse the strategy and analyse the particle content
of the fluxes emitted by the accelerated atom in the light of the
weak-values freshly obtained. We show,
in accord with Unruh's original claim \cite{Unruh}
and contrary to more recent
claims \cite{Grove} \cite{RSG} \cite{mpbrsg},
that, despite its being in thermal equilibrium with the Minkowski
fluctuations, the two level atom emits Minkowski quanta \cite{Unruh92}.
The rate of
production of Minkowski quanta
equals the
rate of internal transitions of the atom.
Note that this production of quanta would also be present in the black hole
case
 if one puts a ''fiducial" \cite{sus1}
detector
(at fixed radius)
in the vicinity of the horizon. The back reaction of the
emitted quanta onto the hole cannot be neglected in view of the very
high temperatures encountered.

In chapter 6 the results obtained in the uniformly accelerated case are
easily mapped to the black hole problem. It is shown how a Hawking photon
emerges from a vacuum fluctuation initially carrying no energy and located
around the light ray that generates the horizon. In the time dependent
background geometry of the collapsing star the vacuum fluctuation breaks up
into two pieces, one of which escapes to infinity and gradually acquires
positive energy to become the post-selected
 Hawking photon, its partner travels beyond the
horizon and ends up in the singularity.

In addition
these vacuum fluctuations
very soon become located on cis-planckian distances while carrying
trans-planckian energy densities \cite{Jacobson}.
This fundamental aspect is presented in a separate publication
\cite{EMP} wherein it is argued that a taming process of these
trans-planckian densities is necessary in view of the nonlinearites
of gravitational interactions. A model for this taming, based on a
Hagedorn-type
of transition \cite{sus2}, is also suggested. Nevertheless an essential
part of black hole physics is as yet unknown, to wit whether or not they
radiate and if so what is the emission process. The study of the
back reaction to Hawking radiation is a necessary concomitant to
understanding this problem. We refer the reader to the following
recent publications:  \cite{tHooft} \cite{Verlinde}
 for some considerations along these lines.

\section{ generalised pre- and post-selection, weak measurements}

Pre- and post-selection consists in specifying both the initial and the final
state of a system (denoted by $S$
in the sequel). The approach developed by
Aharonov et al.\cite{aharo} for studying such pre- and post-selected
ensembles
consists in performing at an intermediate time a "weak measurement"
on $S$. In
essence one studies the first order backreaction
 onto an
additional system
taken  by Aharonov et al. to be a measuring device. But the formalism is
more general. In the case of pair production the additional system could
be the external electric or gravitational field which now has to be
described quantum mechanically. Moreover this formalism can be used to
study the self interaction of
the pairs without introducing the additional system. This is because,
when the first order (or weak)  approximation is
valid, the backreaction takes a simple and universal form governed by  a
c-number, the ''weak value" of the operator which controls the interaction.

In this section the formalism of Aharonov et al. is generalised to
post-selections that do not specify completely the initial and/or the
final state
of the system.
Rather one imposes only that they belong to given subspaces of
the Hilbert space of the system ${\cal H}_S$.
 In this formalism the
post selection remains
a rather formal operation. Therefore in the last part of this chapter we
show  how the post-selection may be realised
operationally following
the rules of quantum mechanics by
coupling to $S$
an additional system in a metastable
state   (the "post selector" $PS$) which will make a transition only if the
system is in the required  final state(s).
The weak value of an operator obtained in this manner changes as time goes by
from an asymmetric form to an expectation value, thereby making contact with
more familiar physics.

The system to be studied is in the (pre-selected) state $\ki$ at
time $t_i$ (more generally pre-selected density matrix $\rho_i$).
The unperturbed time
evolution of $S$ can be described by the following density matrix
\begin{equation} \rho_S(t) = U_S(t,t_i)\ki \bi U_S(t_i,t) \label{weaki}
\end{equation}  where $U_S = \exp (- i H_S t)$ is the time evolution operator
for the system $S$. The post-selection at time $t_f$ consists in specifying
that the system belongs to a subspace, ${\cal H}_S^{\p}$, of
${\cal H}_S$.
The probability to find the system in
this subspace at time $t_f$ is \begin{equation}
P_{\Pi_S^{\p}} = Tr_S \Bigl [ \Pi_S^{\p} \rho(t_f) \Bigr] = Tr_S \Bigl[
\Pi_S^{\p} U_S(t_f,t_i)\ki \bi U_S(t_i,t_f) \Bigr] \label{weakii}
\end{equation}   where
$\Pi_S^{\p}$ is the projection operator onto ${\cal H}_S^\prime$.

Following Aharonov et al. we introduce an additional system, called the "weak
detector" ($WD$), coupled to $S$. The backreaction of $S$ onto $WD$ is
considered, subject to the pre- and post-selection just described. The
interaction hamiltonian between $S$ and $WD$ is taken to be of the form
$H_{S-WD} (t) = \e f(t) A_S B_{WD}$ where $\e$ is a coupling constant,
$f(t)$ is a function, $A_S$ and $B_{WD}$ are hermitian operators acting on
$S$ and $WD$ respectively.

To first order in $\e$ (the coupling is weak), the evolution of the coupled
system $S$ and $WD$ is given by
 \begin{eqnarray} \rho(t_f) =& \Bigl (
U_S(t_f,t_i)U_{WD}(t_f,t_i) - i\e \int_{t_i}^{t_f}\! dt \
 U_S(t_f,t)U_{WD}(t_f,t)  f(t) A_S B_{WD} \times \Bigr. &\nonumber\\ & \Bigl.
\times  U_S(t,t_i)U_{WD}(t,t_i) \Bigr) \ki \ket{WD}\bra{WD}\bi \Bigl(
\rm{h.c.} \Bigr)& \label{weakiii}
 \end{eqnarray} where $U_S$ and $U_{WD}$ are
the free evolution operators for $S$ and  $WD$ and  $\ket{WD}$ is the initial
state of  $WD$. Upon post-selecting at $t=t_f$ that $S$ belongs to the
subspace ${\cal H}_S^\prime$ and tracing over the remaining states of the
system $S$, the reduced density matrix describing the $WD$ is obtained. In
the first order approximation in which we are working it takes a very simple
form \begin{eqnarray}
 \rho_{WD}(t_f) &=& Tr_S
 \Bigl [ \Pi_S^\prime \rho(t_f)
\Bigr]\nonumber\\  &\propto& \Bigl( U_{WD}(t_f,t_i) - i\e
\int_{t_i}^{t_f}\! dt \ U_{WD}(t_f,t)  f(t) A_{S weak}(t)B_{WD}U_{WD}(t,t_i)
\Bigr)\times\nonumber\\ &&\quad\quad\times\ket{WD}\bra{WD} \Bigl( \rm{h.c.}
\Bigr) \label{weakiv}
 \end{eqnarray} where
  \begin{equation} A_{S weak}(t) = {Tr_S
\Bigl [ \Pi_S^\prime U_S(t_f,t) A_S U_S(t,t_i) \ki\bi U_S(t_i,t_f) \Bigr]
\over Tr_S \Bigl [ \Pi_S^\prime U_S(t_f,t_i)\ki\bi U_S(t_i,t_f) \Bigr]}
\label{weakv}
\end{equation}
 is the weak value of $A$. If
one specifies completely the final state, $\Pi_S^\prime =
\ket{\psi_f}\bra{\psi_f}$  then
the result of Aharonov et al.
obtains:
\begin{equation} A_{S weak}(t) = {\bra{\psi_f}
U_S(t_f,t) A_S U_S(t,t_i) \ki  \over \bra{\psi_f} U_S(t_f,t_i) \ki}
\label{weakvi}
 \end{equation}
The remarkable feature of the above formalism is its independence on the
internal structure of the $WD$. The first order backreaction of $S$
onto $WD$ is
universal: it is always controlled by the c-number $A_{S weak}(t)$,
the ''weak
value of $A$". Therefore
if $S$ is coupled to itself by an
interaction hamiltonian, the backreaction
will be controlled by the
weak value of $H_{\rm int}$
in first order perturbation theory.
For instance the modification of the probability that the
final state belongs to ${\cal H}_S^{\p}$ is
given by the imaginary part
of $H_{{\rm int}\ weak}$.

The weak value of $A$ is complex. By  performing a
series of
measurements on $WD$
and by varying the coupling function $f(t)$, the real and imaginary part of
$A_{S weak}$ could in principle be determined. Here the word ''measurement"
must be understood in its usual quantum sense: the average over repeated
realisations of the same situation. This means that the weak value of $A_S$
should also be understood as an average. The fluctuations around $A_{S weak}$
are encoded in the second order terms of eq. \ref{weakiii} which have been
neglected.

To illustrate the role of the real and imaginary parts of $A_{S weak}$, we
recall the example of Aharonov et al consisting of a  weak
detector which has one degree of freedom $q$,
 with a gaussian initial state
$\scal{q}
{WD}=e^{-q^2/
2\Delta^2}$,$-\infty < q < + \infty$.
 The unperturbed hamiltonian of $WD$ is taken to vanish
(hence $U_{WD}(t_1,t_2) = 1$)
 and the interaction hamiltonian is  $H_{S-WD} (t) =\epsilon
\delta(t-t_0) p A_S$ where $p$ is the momentum conjugate to $q$. Then after
the post-selection the state of the $WD$ is given to first order by
\begin{eqnarray}
\scal{q}{WD(t_f)} &=&\left(1-i\epsilon pA_{S weak}(t_0) \right)
e^{-q^2/ 2\Delta^2}\nonumber\\ &=& e^{-i\epsilon pA_{S weak}(t_0)} e^{-q^2/
2\Delta^2}\nonumber\\ &=&e^{-\left(q- \epsilon A_{S weak}(t_0)\right)^2/
2\Delta^2}\nonumber\\ &=& e^{-\left(q- \epsilon {\rm Re }  A_{S
weak}(t_0)\right)^2/ 2\Delta^2} e^{+i\epsilon q {\rm Im}A_{S weak}(t_0)/
\Delta^2} \label{weakviii}
 \end{eqnarray}
 The real part of $A_{S weak}$ induces a
translation of the centre of the gaussian, the imaginary part
a change in the
momentum. Their effect on the $WD$ is therefore measurable.
The validity of the first order approximation requires
$\e A_{S weak} / \Delta <<1$.

It is instructive to see how unitarity is realised in the above formalism.
Take
 $\Pi^i_S$
to be  a
complete orthogonal set of projectors acting on the
Hilbert space of $S$. Denote by $P_i$ the probability that
the final state of the system belong to the subspace spanned by $\Pi^i_S$ and
by $A_{S weak}^i$ the corresponding weak value of $A$. Then the mean value of
$A_S$ is  \begin{equation} \bi A_S \ki =
\sum_i P_i A_{S weak}^i  \label{weakix} \end{equation} Thus the mean
backreaction if no post-selection is performed is the average over the
post-selected backreactions. Notice that the imaginary parts of the weak
values necessarily cancel since the l.h.s. of eq. \ref{weakix} is
real.

Up to now the postselection has been implemented by projecting
by
hand
the state of
the system onto a certain subspace ${\cal H}^{\p}_S$.
Such a projection may be
realised operationally by introducing
an additional
quantum
system, a ''post-selector" ($PS$),
coupled in such a way
 that it will make a transition if and only if the system $S$
is in the required final state. Then by considering
only that subspace of the
final states in which $PS$ has made the
transition, the pre- and
post-selected ensemble is recovered. This
quantum description of the
post-selection
will turn
out to be very useful when considering individual Hawking quanta.
The detected
Hawking photons will then be analysed using the weak measurement
formalism.

We shall consider the very simple model of a $PS$ having two states,
initially in the ground state, and coupled to the system by an interaction
of the form \begin{equation}
H_{S-PS} = \la g(t) ( a^\dagger Q_S + a Q_S^\dagger)
\label{weakx}
\end{equation}
where $\la$ is a
coupling constant, $g(t)$ a time dependent function, $a^\dagger$  the operator
that induce transitions from the ground state to the exited state of the
$PS$ , $Q_S$ an operator acting on the system $S$. The
postselection is performed at $t=t_f$ and consists in finding the $PS$ in the
exited state.

For simplicity we shall work to second order in $\la$ (although in principle
the interaction of $PS$ with $S$ need not be weak).
The wave function of the combined system $S+WD+PS$ is in interaction
representation
\begin{eqnarray}
&&\left[
1 -i \int dt \left( H_{S-WD}(t) + H_{S-PS}(t) \right)\right.
\nonumber\\
&&
\quad\left.
-{1 \over 2}
\int dt \int dt^\prime
{\cal T}\left[ H_{S-WD}(t) + H_{S-PS}(t) \right)
\left( H_{S-WD}(t^\prime) + H_{S-PS}(t^\prime) \right]\right]\nonumber\\
&&\quad\quad\ket{\psi_i}\ket{WD}\ket{0_{PS}}
\label{weakxi}\end{eqnarray}
where $\ket{0_{PS}}$ is the ground state of $PS$ and ${\cal T}$ is
the time ordering operator. Upon imposing that the
$PS$ be in its excited state at $t=t_f$ the
resulting
wave function of $S$ and $WD$ is
, to order $\epsilon$,
\begin{eqnarray}
&&\quad \left [
-i \int dt \la g(t) Q_S(t)  \right.
\nonumber\\
&& \left. -
\int dt \int dt^\prime
{\cal T}
\left[\epsilon f(t) A_S(t) B_{WD}(t) \la g(t^\prime) Q_S(t^\prime)
\right]
\right]
\ket{\psi_i}\ket{WD}
\label{weakxii}\end{eqnarray}
Then
tracing over the states of $S$
yields the reduced density matrix of $WD$
\begin{equation}
\left[
1 - i \epsilon \int dt_0 f(t_0) B_{WD}(t_0)
A_{S weak}(t_0) \right] \ket{WD}\bra{WD} \left[ {\rm h.c.} \right]
\label{weakxiii}
\end{equation}
where
\begin{equation} A_{S weak}(t_0) =
{ \bi \int dt g(t) Q_S^\dagger (t)  \int dt^\p
g(t^\p) {\cal T } \left[ A_S(t_0) Q_S(t^\p) \right]
\ki \over \bi \int dt g(t) Q_S^\dagger (t) \int
dt^\p g(t^\p) Q_S(t^\p) \ki} \label{weakxiv} \end{equation}
Note that the weak value of $A_S$ results from the interference of the two
states in eq. \ref{weakxii}.

There are
several cases when eq. \ref{weakxiv} takes a simple form.
If $g(t)$ is non vanishing only after $t=t_0$ the time ordering operator is
very simple to implement and the resulting expression for $A_W$ takes a
typical (for a weak value) asymmetric form \begin{equation}
A_{S weak}(t_0) =  {
\bi \int dt g(t) Q_S^\dagger (t)
 \int dt^\p g(t^\p) Q_S(t^\p)  A_S(t_0) \ki \over \bi \int dt g(t) Q_S^\dagger
(t) \int dt^\p g(t^\p) Q_S(t^\p) \ki} \label{weakxv} \end{equation}
If in addition $g(t)=\delta(t-t_f)$, $t_f >t_0$ and $Q_S=\Pi_S^\p$,
eq. \ref{weakv} is recovered since $ (\Pi_S^\p)^2 = \Pi_S^\p$.

If on the other hand $g(t)$ is non vanishing only before $t=t_0$ then
the time
ordering operator becomes trivial once more and eq. \ref{weakxiv} takes the
form
\begin{equation} A_{S weak}(t_0) =  { \bi \int dt g(t) Q_S^\dagger
(t) A_S(t_0)
 \int dt^\p g(t^\p) Q_S(t^\p)   \ki \over \bi \int dt g(t) Q_S^\dagger (t)
\int dt^\p g(t^\p) Q_S(t^\p) \ki} \label{weakxvi}
\end{equation}
This is the
expectation value of $A_S$ if the $PS$ has made a transition. It is
necessarily real.
The weak value of $A_S$ if the $PS$ has not made a
transition can also be computed. It is related to the mean value of $A_S$
and to eq. \ref{weakxiv} through the unitary relation eq. \ref{weakix}.

In the sequel we shall use the above formalism to compute the
weak value of $T_{\mu\nu}$. The
justification for considering the weak value follows from the abovementioned
universality, which implies that it is the source of the first
order backreaction.

\section{Post-selecting Rindler quanta and the weak value of $T_{\mu\nu}$}

\subsection{Introduction}

In this chapter we shall post-select the presence of Rindler quanta in
Minkowski vacuum. The weak value of the energy momentum tensor is obtained in
this pre- and post-selected ensemble.

This program is
carried out in a mechanical way by
post-selecting the presence of a pair of Rindler particles of fixed Rindler
energy $\la$ in Rindler vacuum. This is a straightforward calculation but it
presents several unsatisfactory aspects:   \begin{description}
\item[1] The
post-selected
state is specified in both the right and left Rindler quadrant
whereas in the analogous black hole problem only the region outside the
horizon is accessible to measurements.
\item[2] The state that is
post-selected has probability zero of being realised  \linebreak[4]
($\scal{{\rm pair\ of\ rindlerons}}{0_M} \equiv 0$ ) even though the weak
value of $T_{\mu \nu}$ is finite (it is a conditional measurement: a
cancellation of zeros occurs between the numerator and denominator of eq.
\ref{threefouri}).
\item[3] Post-selection is a formal procedure. Physical insight would be
gained by introducing
an additional system (the "post-selector" of
chapter 2)
 that gets correlated to the field thereby realising
operationally
the post-selection.  \item[4] The energy momentum tensor
obtained is singular on the horizons. As shown in ref \cite{tmunu},
the appearance  of
such singularities is a generic feature when working with Rindler
modes.
\end{description}
Most of this
chapter
and the next one will be devoted to
solving these problems.

The first one finds its solution in the partial post-selections
introduced  in
section 2. The state of the field is post-selected only in the right Rindler
quadrant while tracing over the state in the left
one.

The second problem is solved by performing an even less restrictive
post-selection.
One post-selects the presence of one
Rindleron in the mode of interest in the right Rindler quadrant while tracing
over the other
modes in that quadrant
and over the entire state of the field in the left quadrant.

In order to
reveal the physical meaning of post-selection we shall work in
the next chapter in a more physical set-up (solving 3). We
 shall consider a uniformly accelerated two level atom in
Minkowski vacuum initially in its ground state, coupled to the field during a
proper time ${T}$. The post-selection shall consist in finding the two
level atom in its ground or its excited state.
Since
the two level atom interacts with the field for a finite time it
effectively post-selects the presence of a wave packet rather than a
delocalised mode.
We shall show that if the post-selected wave packet is
sufficiently tight then the energy momentum tensor will be regular
everywhere. The singularities mentioned in 4 have disappeared.

We shall restrict our analysis to a massless scalar field in 1+1
dimensions in view of its simplicity and of its relation to the emission
of
quanta
by a black hole. The simplicity arises from the conformal invariance
which implies also some very particular properties.
 For instance
the propagator is infrared divergent
and is the sum of a left and a right
moving part. Hence the energy momentum tensor takes
the form $T_{UU}=f(U)$, $T_{VV}=g(V)$,
$T_{UV}=0$ where $V,U = t \pm x$ are the Minkowski light like coordinates.
These properties will result most notably in the {\it eternal} life
of vacuum fluctuations. This is to be contrasted with the vacuum fluctuations
of a massive field which exist only in a finite region of space time
(typically $\Delta t \simeq \Delta x \simeq m^{-1}$).
 We thus expect that any breaking of the
conformal invariance, a small mass or
an interaction with another field, will
change dramatically the structure at large distances of the vacuum
fluctuations we shall exhibit. The same reservations apply to the
black hole problem as well.

\subsection{The kinematics of a massless scalar field in 1+1 dimensions}

In this section, we review the relevant properties of the Rindler
quantisation of the scalar field.
The conformal invariance of this massless scalar field is best
exploited by using the light like coordinates $U,V$ introduced above.
Whereupon the Klein-Gordon equation takes the form $\partial_U\partial_V \phi
= 0$ and any solution can be written as
\begin{equation} \phi(U,V) = \phi(U) + \phi(V) \label{threethreei}
\end{equation} From now on we shall drop the right moving piece and consider
the ''$V$" term only. It is
obvious that all conclusions shall be equally valid
for
the
right
movers.

The second quantised field can be decomposed into a basis of
Minkowski modes
\begin{equation} \phi(V) = \int_0^{\infty} d\omega  \left( a_{\om}
\varphi_{\om} (V) + a_{\om}^\dagger \varphi_{\om}^* (V)\right)
\label{threethreeii} \end{equation} \begin{equation} \varphi_{\om} (V) = {
e^{-i\om V} \over \sqrt{4 \pi \om}} \label{threethreeiii} \end{equation} The
Minkowski vacuum $\ket{0_M}$ is that annihilated by all the $a_{\om}$'s. The
propagator in Minkowski vacuum is \begin{equation} G_+(V,V^\prime)
=\expect{0_M}{ \phi(V)\phi(V^\prime)}{0_M} = -{1 \over 4 \pi} \log (V-V^\prime
- i \e) \label{threethreeiv} \end{equation} The (normal ordered)
hamiltonian of the field is
(for left movers) \begin{equation} H_M= \int_{-\infty}^{+\infty} \! dV T_{VV}
=  \int_0^{+\infty} \! d\om {\om }  (
a_{\om}^\dagger  a_{\om} ) \label{threethreev} \end{equation}

The uniformly accelerated observer
will be
taken to be in the right
Rindler
quadrant $U<0,V>0$. In this quadrant one defines Rindler coordinates $\rho ,
\tau$ by \begin{equation} \left \{ \begin{array}{ll} t=\rho\ {\rm sinh} a\tau
\\ x=\rho\ {\rm cosh} a\tau \end{array} \right. \label{threethreevbis}
\end{equation} and Rindler light like coordinates $u,v$ by \begin{eqnarray}
&u,v = \tau \mp a^{-1} \ln \rho a  \nonumber\\ &\left \{ \begin{array}{ll}
U=-a^{-1}\ e^{-au} \\ V=a^{-1}\ e^{av} \end{array} \right.
\label{threethreevi} \end{eqnarray} where $a$ is the acceleration. These
coordinates may be extended to the left Rindler quadrant by the analytic
continuation $\tau \rightarrow \tau \pm i \pi/a$.

The natural basis of quantisation a uniformly accelerated observer would
choose  is the Rindler basis which consists of plane waves in the variables
$u,v$ (Rindler modes). But the Bogoljubov transformation
from the
  Minkowski
modes to the Rindler modes is singular \cite{tmunu}
 and care must be
taken to define it as a limit if the Minkowski properties of the theory are
to be satisfied.
To this end it is useful to first introduce an alternative basis
of Minkowski modes \cite{Unruh}:
\begin{eqnarray}  \varphi_{\la,M}(V) &=& \int_0^{\infty}\! d\omega\
\gamma_{\la,\om} \varphi_{\om}(V) \nonumber\\ &=& {\left[ a(\e +
iV)\right]^{-i \la /a} \over  \sqrt{ ( e^{\pi \la /a} - e^{-\pi \la /a})4 \pi
\la}}\nonumber\\  &\simeq& {e^{\pi \la/2a}\over  \sqrt{\vert e^{\pi \la /a} -
e^{-\pi \la /a}\vert}} { (aV)^{-i \la /a} \over \sqrt{4 \pi \vert \la\vert}} +
{e^{-\pi \la/2a}\over \sqrt{\vert e^{\pi \la /a} - e^{-\pi \la /a}\vert}}
{\vert aV \vert^{-i \la /a} \over \sqrt{ 4 \pi \vert \la\vert}}
\label{threethreevii} \end{eqnarray}

where \begin{equation}  \gamma_{\la,\om}=
\left({1\over \Gamma({i\la\over a})}  \sqrt{ \pi \over {\la \over a} {\rm
sinh}{\pi \la \over a} }\right) {1 \over \sqrt{2 \pi a \om}} \left( {\om \over
a } \right)^{i\la\over a} e^{-\om \e} \label{threethreeviii} \end{equation}

The first factor in $\gamma_{\la,\om}$ is a pure phase introduced for
convenience. The factor $ e^{-\om \e}$ is the crux of the construction.
It
defines the integral eq. \ref{threethreevii} and ensures the correct
Minkowski
properties of the theory. For instance it gives the correct pole
prescription
at $V=V^\prime$ of the propagator eq. \ref{threethreeiv} when expressed in
terms
of the modes $\varphi_{\la,M}$ as $G_+(V,V^\prime) = \int_{-\infty}^{+
\infty}
d \la   \varphi_{\la,M}(V)\varphi_{\la,M}^*(V^\prime)$. The limit $\e
\rightarrow 0$ is to be taken at the end of all calculations.

The  annihilation and creation operators corresponding to
the modes $ \varphi_{\la,M}$ are $a_{\la,M}$ and $a_{\la,M}^\dagger $.

The right and left Rindler modes  $\varphi_{\la,R} (V)$ and $\varphi_{\la,L}
(V)$ are now defined by the Bogoljubov
transformation
\begin{equation}
\left\{
\begin{array}{ll} \varphi_{\la , M} = \alpha_{\la} \varphi_{\la,R} +
\beta_{\la} \varphi_{\la,L}^* \\ \varphi_{- \la , M} = \beta_{\la}
\varphi_{\la,R}^* + \alpha_{\la} \varphi_{\la,L} \label{threethreeix}
\end{array} \right. \quad \la>0 \label{threethreex} \end{equation}
with
\begin{equation}  \alpha_{\la} = { e^{\pi \la / 2 a} \over \sqrt { e^{\pi \la
/ a} - e^{- \pi \la / a}}} \quad \beta_{\la} =  { e^{-\pi \la / 2 a} \over
\sqrt { e^{\pi \la / a} - e^{- \pi \la / a}}} \label{threethreexi}
\end{equation} In the limit $\e \rightarrow 0$ the
Rindler modes take the familiar form
\begin{equation} \begin{array}{ll} \varphi_{\la,R} (V) =
\theta(V){1 \over \sqrt{4 \pi \la}}{
(aV)^{-i\lambda/a}
} =
{1 \over \sqrt{4 \pi \la}}  e^{-i\la v} \\ \varphi_{\la,L} (V) =
\theta(-V) {1 \over \sqrt{4 \pi \la}} \vert a V
\vert ^{-i\lambda/a}
\end{array}
\label{threethreexii} \end{equation} For finite $\e$ they differ from these
limiting forms only
when $V \leq \e$.
To these modes are associated the
Rindler destruction and creation operators $a_{\la,R},
a_{\la,R}^\dagger $ and $a_{\la,L}, a_{\la,L}^\dagger $.

Using the above Bogoljubov transformation it is easy to show that
\begin{equation} \ket{0_M} = \prod_\la {1 \over \alpha_\la} e^{-{ \beta_\la
\over \alpha_\la} a^\dagger_{\la,L} a^\dagger_{\la,R}} \ket{0_{RL}}
\label{threethreexiii} \end{equation}
where $\ket{0_{RL}}= \ket{0_{R}} \otimes\ket{0_{L}}$ is Rindler vacuum in
both the right(R) and left(L) quadrants. Upon tracing over the left quanta the
reduced density matrix in the right Rindler quadrant is an exact thermal
distribution of right rindlerons. This proves that Rindler physics in
Minkowski vacuum is identical to working
in a thermal bath at temperature
${a/2 \pi}$.

\subsection{Post-selecting Rindler quanta}

 We are now ready to evaluate the weak value of $\phi(V)\phi(V^\prime)$
when the
pre-selected state is Minkowski vacuum and the post-selected state contains
Rindler quanta. The energy momentum tensor is
then  readily obtained by the limiting process  $T_{VV} = \lim_{V^\prime
\rightarrow V} \partial_V \phi(V)\partial_{V^\prime}\phi(V^\prime)$.
The first post-selected state considered is a pair
of Rindlerons in Rindler vacuum.
By availing
oneself of the identity
\begin{equation}
  a_{\la,R}^\dagger a_{\la^\prime,L}^\dagger \ket{0_{RL}} ={1 \over \al_\la
\al_{\la^\prime} } a_{\la,M}^\dagger a_{-\la^\prime,M}^\dagger
\ket{0_{RL}} + {
\be_\la \over \al_\la} \delta ( \la -
\la^\prime )\ket{0_{RL}} \end{equation}
it is
straightforward to obtain the weak value of $\phi(V)\phi(V^\prime)$

\begin{eqnarray} &{ \expect
{0_{RL}}{a_{\la,R}a_{\la,L} \phi(V)\phi(V^\prime)}{0_M} \over  \expect
{0_{RL}}{a_{\la,R}a_{\la,L}}{0_M} } \quad\quad\nonumber\\ &= {2 \over
\alpha_{\la}\beta_{\la}} \varphi_{-\la , M}^* (V)
 \varphi_{\la , M}^* (V^\prime) \ + \  {
\expect{0_{RL}}{\phi(V)\phi(V^\prime)}{0_M} \over  \scal {0_{RL}}{0_M} }
\label{threefouri} \end{eqnarray}
 It decomposes into two terms. The first depends on the quantum number $\la$
and is the contribution of the post-selected pair of rindlerons.
It carries an energy density equals  \begin{equation} \lim_{V^\prime
\rightarrow
V}
\partial_V \partial_{V^\prime}{2 \over \alpha_{\la}\beta_{\la}} \varphi_{-\la
, M}^* (V)
 \varphi_{\la , M}^* (V^\prime)= {\la \over 2 \pi a^2}{1 \over (V + i \e)^2}
\label{threefourii} \end{equation}

The second term is independent of $\la$ and appears because except
for the mode
$\la$, Rindler vacuum has been post-selected (if
Rindler vacuum is post-selected only the second term appears). It is
convenient to rewrite it as the sum of the expectation value of $T_{VV}$ in
Minkowski vacuum\footnote{
In this section and the following one we shall explicitly write the vacuum
expectation value of the energy momentum tensor $\expect{0_M}{T_{VV}}{0_M}$
even though it vanishes. It is kept only to facilitate the transcription of
these results  to the black hole problem
where the vacuum expectation of the energy is non trivial and must be
renormalised
carefully.} plus another term \begin{eqnarray} &&{
\expect{0_{RL}}{T_{VV}}{0_M} \over \scal {0_{RL}}{0_M} }\nonumber\\  &=&
\lim_{V^\prime \rightarrow V} \partial_V\partial_{V^\prime} \int_0^\infty \!
d\la\  -2 {\beta_{\la} \over \alpha_{\la} }\varphi_{-\la,M}^*(V)
\varphi_{\la,M}^*(V^\prime) \ + \ \expect{0_M}{T_{VV}}{0_M}\nonumber\\
&=&-{\pi
\over 12} \left( a \over 2 \pi \right)^2 {1 \over a^2(V + i \e)^2} \ + \
\expect{0_M}{T_{VV}}{0_M} \label{threefouriii} \end{eqnarray}

These results allow for two complementary interpretations. The Rindler
point of view, restricted to $V>0$, is obtained by considering the
Rindler energy density $T_{vv} = \left(
d V / dv \right)^2 T_{VV}$.
Then eq. \ref{threefourii} gives  the energy density of the post-selected
 rindleron $\la/2 \pi$, the jacobian being
$\left( d V / dv \right)^2 = a^2 V^2$.
Since we have imposed that no other Rindleron be present in the final
state, the remaining term eq. \ref{threefouriii} yields
the value of the Rindler vacuum energy which is minus the thermal energy
density at a temperature $a/2 \pi$ (Minkowski vacuum
contains a thermal distribution of rindlerons).
The Minkowski point of view is completely different.
Since the hamiltonian
 $H_M$ is diagonal in
$\omega $
and annihilates
Minkowski vacuum
 $H_M |0_M>=0$, eq. \ref{threefourii} and  eq. \ref{threefouriii} contain
zero Minkowski energy. Indeed, the pole prescription at the
horizon $V=0$ ensures that their integrals over the entire
domain of $V$ vanish.

We note that the energy in the left quadrant is identical to
that in the right quadrant since there is a complete symmetry between the two.
This symmetry is manifest
by considering the total Rindler energy (the boost operator)
\begin{equation} H_R = \int_{-\infty}^{+\infty}\!\! dv T_{vv}({\rm right}) \ -
\   \int_{-\infty}^{+\infty}\!\! dv T_{vv}({\rm left}) = \int_{0}^{+\infty}\!
d\la \la (a_{\la,M}^\dagger a_{\la,M} - a_{-\la,M}^\dagger a_{-\la,M})
\label{threefouriv} \end{equation} The minus signs arise because Rindler time
runs backwards in the left Rindler quadrant. Since Minkowski vacuum is an
eigenstate of $H_R$ (it is invariant under boosts) with eigenvalue zero the
total Rindler energy in the pre- and post-selected ensemble considered above
must vanish.
Hence, the Rindler energy
in the right quadrant
is equal
and opposite
 to the energy in the left quadrant.

The post-selection that was used in the calculation above was rather crude and
we shall now refine it in three successive steps.

We first consider
post-selections that are performed
only in the right quadrant while tracing over the state of the field in the
left quadrant. Post-selecting one rindleron of energy $\la$ in the right
quadrant is achieved using the projector  \begin{equation} \Pi_{\la,R}
=  I_{L} \otimes
a^\dagger_{\la,R}  \ket{0_{R}}\bra{0_{R}} a_{\la,R}
 \label{threefourv} \end{equation}
where $I_{{L (R)}}$ is the identity operator restricted to the left (right)
quadrant and $\ket{0_{L (R)}}$ is Rindler vacuum in the left (right)
quadrant. Using the formalism developed in the preceding chapter the
corresponding weak value of $T_{VV}$ is given by \begin{equation} { \expect
{0_M}{\Pi_{\la,R} T_{VV}}{0_M} \over  \expect {0_M}{\Pi_{\la,R}}{0_M} }
\label{threefourvi} \end{equation} It leads back
 to eq. \ref{threefouri} because of the EPR correlation's between the two
quadrants: if there is a rindleron on the right then their
necessarily also is a rindleron on
the left (its partner) with the
opposite
 Rindler
energy. This partenaria follows from eq. \ref{threethreexiii} where
the operators
$a^\dagger_{\la,R}$ and $a^\dagger_{\la,L}$
appear
 in product only.
 In the
black hole problem
the equivalent
EPR correlation's will mean that
 to each outgoing Hawking photon their
corresponds an ingoing partner on the other side of the horizon.

An even less restrictive post-selection consists in specifying only partially
the state of the field in the right quadrant. One chooses
that the final state contains one rindleron on the right in the mode $\la$
while tracing over all other right rindlerons and over all left rindlerons.
The resulting projector is
\begin{equation} \tilde \Pi_{\la,R}  =
I_{L} \otimes \prod_{\la^\prime \neq \la}
I_{\la^\prime,R}\otimes \ket{1_{\la,R}}\bra{1_{\la,R}}  \label{threefourvii}
\end{equation}
where $I_{\la,L(R)}$ is the identity operator restricted to the
mode $\la$ in the left (right) quadrant and $\ket{1_{\la, R}}$ is
the one particle state restricted to the mode $\la$.
The
corresponding weak value of $T_{VV}$ is \begin{eqnarray} {
\expect{0_M}{\tilde \Pi_{\la,R} T_{VV}}{0_M} \over \expect{0_M}{\tilde
\Pi_{\la,R}}{0_M}} =
{\la \over 2 \pi
a^2}{1 \over (V + i \e)^2} \ + \  \expect{0_M}{T_{VV}}{0_M}
\label{threefourviii} \end{eqnarray}
The first term is the energy of the Rindleron $\la$ already obtained
in eq. \ref{threefouri} and eq. \ref{threefourvi}. The second term
is simply the Minkowski vacuum expectation value since no further
post-specification is imposed on the final state.
This is why the probability $\expect{0_M}{\tilde
\Pi_{\la,R}}{0_M}$ to be in the eigenspace of $\tilde
\Pi_{\la,R}$ is finite. This is to be opposed to the probabilities
encountered previously (the denominators of eq. \ref{threefouri} and
eq. \ref{threefourvi} ) which
vanish because all the Rindler modes have been
specified
to be in their Rindler ground state.
 In physically realistic situations such as considered in the next
chapter only
nonvanishing probabilities
will occur since a finite number of modes will be coupled to
the "post-selector".
Nevertheless the weak values eq. \ref{threefourvi}
and eq. \ref{threefourviii} can be formally related by a unitary
relation similar to eq. \ref{weakix}: by taking a set
of orthogonal projectors like $\Pi_{\la,R}$ whose combined eigenspace
is equal to the eigenspace of
$\tilde \Pi_{\la,R}$
and summing
the corresponding
weak values multiplied by the relative probabilities that they occur,
eq. \ref{threefourviii} is recovered. In order to realise this unitary relation
one must post-select
the presence of two, three, any number of rindlerons.
The induced
weak values of $T_{VV}$ are easily obtained and the contribution of each
individual post-selected particle is found to be independent
(if the particles are orthogonal) of the post-selection performed
on the other particles. In other words, for a free field the vacuum
fluctuations of orthogonal particles are independent of each other.

We finally consider the post-selection of wave packets. Instead of the
projector eq. \ref{threefourv}, we define:
\begin{equation} \Pi_{v_0,\la_0,R} = I_L \otimes a^\dagger_{v_0,\la_0,R}
\ket{0_R
} \bra{0_R} a_{v_0,\la_0,R} \end{equation} where
$a_{v_0,\la_0,R} = \int_0^{+\infty}\! d\la f(\la) a_{\la,R}$ is the
destruction operator of a wave packet of right rindlerons centred around
$v=v_0$ and $\la=\la_0$. The state $\Pi_{v_0,\la_0,R} \ket{0_M} $ is
 \begin{equation} \Pi_{v_0,\la_0,R} \ket{0_M}
= \left( \int_0^{+\infty}\! d\la f^*(\la) a^\dagger_{\la,R} \right)
\left( \int_0^{+\infty}\! d\la^\prime -
{\be_{\la^\prime} \over \al_{\la^\prime}}
f(\la^\prime) a^\dagger_{\la^\prime,L} \right)  \ket{0_{RL}}
\end{equation} where the EPR correlated wave packet in the left Rindler
quadrant appears explicitly.
Note the dissymetry of the wave packets: the
induced
wave packet
in the
left quadrant contains the factor $\be_{\la^\prime} / \al_{\la^\prime}$
since it originates
from
 the EPR correlations in eq. \ref{threethreexiii}. This dissymetry will play
a fundamental role when analysing the flux emitted by the accelerated
detector and the black hole.
The weak value of $T_{VV}$ is
\begin{eqnarray} {
\expect{0_M}{ \Pi_{v_0,\la_0,R} T_{VV} }{0_M}
\over
\expect{0_M} { \Pi_{v_0,\la_0,R} } {0_M}
}  = &
2 \left[
 \int_0^{\infty}\! d\la \int_0^{\infty}\! d\la^{\p}
{\be_{\la^\prime} \over \al_\la \al_{\la^{\prime}}^2 }
f^*(\la) f(\la^{\p}) \partial_V
\varphi_{\la,M}^* \partial_V \varphi_{-\la^{\p},M}^* \right]
 \nonumber\\ &
\times\ \left[  \int_0^{+\infty}\! d\la
{ \be_\la^2  \over \al_\la^2} \vert f(\la) \vert^2
 \right] ^{-1}
+ {
\expect{0_{RL}} {  T_{VV} }{0_M} \over \scal{0_{RL}}{0_M} }
\label{threefourx}
\end{eqnarray}
Notice that eq. \ref{threefourx}
is complex due to the presence of the factor $\be_{\la^\prime}
/ \al_{\la^\prime}$.

\section{The uniformly accelerated two level atom.}

\subsection{Post-selection of wave packets by a two level atom.}

The set-up consists of a two level atom initially in its ground
state at $t=-\infty$, coupled to the field during a finite time.
The post-selection is realised
by imposing that the atom be in its ground or its excited state at
$t=+\infty$. In order to reveal the structure of the vacuum fluctuation
which induces the transition, we shall consider the weak value of
$T_{\mu\nu}$. As discussed in chapter 2 these weak values would be the
source of the linear gravitational backreaction. They could also be measured
by a weak detector.

In this section, we consider for generality that the two level atom
follows an arbitrary
trajectory: $t=t(\tau),x=x(\tau)$ where $\tau$ is the proper time along the
trajectory.
The hamiltonian of the two level atom is given by   \begin{equation}
\int\! dt dx\ H_{\rm atom} (t,x) = \int\! d\tau\  \left\{
m A^\dagger (\tau)A(\tau)
 + g m \left[
f(\tau) \phi(t(\tau),x(\tau)) A(\tau) + {\rm h.c.} \right]
\right\}  \label{threefivei}
\end{equation} where $g$ is a dimensionless coupling constant that shall be
taken for simplicity  small enough that second order perturbation theory be
valid, $m$ is the difference of energy between the ground and the
excited state
of the atom,
$A(\tau)=A e^{-im\tau}$ is the operator that induces a transition
 from the excited state to the
ground state of the atom
and $f(\tau)$ is a function that governs when the interaction
is turned on and off.
In interaction representation eq. \ref{threefivei}
becomes      \begin{eqnarray} \int\! dt dx\ H_{\rm int} (t,x) &=& g m \int\!
d\tau \left[ e^{-im\tau} f(\tau) \phi(t(\tau),x(\tau)) A + {\rm h.c.}
\right]\nonumber\\ &=& g m  \left[ \phi_m^\dagger A + {\rm h.c.} \right]
\label{threefiveii} \end{eqnarray} where we have introduced for convenience
the field operator \begin{equation} \phi_m = \int\! d\tau e^{+im\tau}
f^*(\tau) \phi(\tau) \label{threefiveiii} \end{equation} and its
hermitian conjugate $\phi_m^\dagger$.

The probability $P_e$ for the two level atom to get excited is, in second
order perturbation theory
  \begin{eqnarray}
P_e&=& P_{e,v} + P_{e,u} = 2 P_{e,v}\nonumber\\
P_{e,v}&=& g^2 m^2 \expect{0_M}{
\int\! d\tau e^{-im\tau} f(\tau) \phi(\tau) \int\! d\tau^{\p}
e^{+im\tau^{\p}} f^*(\tau^{\p}) \phi(\tau^{\p}) } {0_M} \nonumber\\ &=&
g^2 m^2 \expect{0_M}{\phi_m^\dagger  \phi_m}{0_M} \label{threefiveiv}
\end{eqnarray}
where $P_{e,v}$ and $P_{e,u}$ are the contribution to the probability
of left movers and right movers.

If $e^{-im\tau} f(\tau)$ contains no negative frequencies in its Fourier
transform with respect to $\tau$ then eq. \ref{threefivei}
defines
 a Lee model:
when it is inertial it only responds to the presence
of real particles. When following a non inertial trajectory it responds to the
presence of local quanta. For a Lee model $f(\tau)$ may not decrease as
quickly as an exponential when $\tau \rightarrow \pm \infty$. This constraint
will be seen to be very strong and we shall work necessarily with
non Lee models which can spontaneously
excite.
However by
choosing $f(\tau)$ such that the negative frequency part of $e^{-im\tau}
f(\tau)$ is exponentially small this spontaneous  excitation is
exponentially small as well.

 We now consider the weak values of $T_{\mu\nu}$. As described in chapter 2
they take the form
\begin{equation} \bk{T_{\mu\nu}(t_0,x_0)}_{weak\  e} =
{g^2 m^2  \over 2 P_{e,v} }
\expect{0_M}{\phi_m^\dagger  {\cal T}\left[ T_{\mu\nu}(t_0,x_0) \int\! d\tau
e^{+im\tau} f^*(\tau) \phi(\tau) \right] }{0_M}
 \label{threefivev} \end{equation}
where ${\cal
T}$ is the time ordering operator and the subscript $e$ refers to the
post-selection of the two level atom in its excited state at $t=+ \infty$.

If the interaction lasts only from $\tau_i$ to $\tau_f$, i.e. $f(\tau)=0$ for
$\tau<\tau_i$ or $\tau>\tau_f$, then in the past of the future light cone
centred on $t(\tau_i),x(\tau_i)$ (this region of space time shall be called
$I_-$) or in the future of the past light cone centred on
$t(\tau_f),x(\tau_f)$ (denoted $I_+$)
the time ordering
operator ${\cal T}$ is trivial to implement and  $\bk{T_{\mu\nu}}_{weak\  e}$
takes
the simple form
 \begin{equation} \bk{T_{\mu\nu}(I_-)}_{weak\  e} =
{g^2 m^2  \over 2 P_{e,v} }
\expect{0_M}{ \phi_m^\dagger  \phi_m T_{\mu\nu}} {0_M}
 \label{threefivevi} \end{equation}
\begin{equation} \bk{T_{\mu\nu}(I_+)}_{weak\  e} =
{g^2 m^2  \over 2 P_{e,v} }
 \expect{0_M}{ \phi_m^\dagger
 T_{\mu\nu} \phi_m} {0_M}
\label{threefivevii} \end{equation}  In the regions where $I_+$ and $I_-$
overlap these two expressions coincide because $T_{\mu\nu}(t_0,x_0)$  and
$\phi_m$ commute since  the point where $T_{\mu\nu}$ is evaluated is seperated
from the trajectory by a space like distance.

The probability that the two level atom is found in its ground state, at
$t=+\infty$, is $P_g =
1-P_e$. When this occurs the weak value of $T_{\mu\nu}$ takes the form
\begin{equation}
\bk{T_{\mu\nu}(I_-)}_{weak\  g} =
{1 \over  P_{g}}
\left[
\expect{0_M}{T_{\mu\nu}}{0_M}
\ - \
  g^2 m^2  \expect{0_M}{ \phi_m^\dagger  \phi_m T_{\mu\nu}} {0_M}
\right]
\label{threefiveviii}
\end{equation}
\begin{eqnarray}
&&\quad\bk{T_{\mu\nu}(I_+)}_{weak\  g} =
{1 \over  P_{g}}
\left[
\expect{0_M}{T_{\mu\nu}}{0_M}
\ - \ \right.\nonumber\\
&& \left. 2 g^2m^2 {\rm Re} \left [ \expect{0_M} {  T_{\mu\nu} \int
\! d\tau_2 \int^{\tau_2} \! d \tau_1 e^{-i m \tau_2} f(\tau_2) \phi(\tau_2)
e^{+i m \tau_1} f^*(\tau_1) \phi(\tau_1) }{0_M} \right]
\right]\quad
\label{threefiveix}
 \end{eqnarray}
where for simplicity we have given only
the expressions valid in $I_-$ and $I_+$. The first term in
eq. \ref{threefiveviii} and eq. \ref{threefiveix} comes from that part of the
wave function
in which the two level atom has
remained for all times in its ground state whereas the second term is an
interference effect between those amplitudes wherein the
atom has
remained
in its ground state and those amplitudes wherein it got
excited and deexcited successively.

The weak values eq. \ref{threefivevi}, eq. \ref{threefivevii} and
eq. \ref{threefiveviii},
eq. \ref{threefiveix} are related to the mean energy momentum by the unitary
relation    \begin{equation} \bk{T_{\mu\nu}} = P_e
\bk{T_{\mu\nu}}_{weak\  e} +
P_g \bk{T_{\mu\nu}}_{weak\  g}  \label{threefivex} \end{equation}
If $(t_0,x_0)$
belongs to $I_-$,  $\bk{T_{\mu\nu}}$  is simply
\begin{equation}  \bk{T_{\mu\nu}(I_-)} = \expect{0_M}{T_{\mu\nu}}{0_M}
  \end{equation}
since the interaction has not yet taken place. On the contrary, if
 $(t_0,x_0)$
belongs to $I_+$,  $\bk{T_{\mu\nu}}$  is the mean energy radiated
\begin{eqnarray}  &&\bk{T_{\mu\nu}(I_+)} =
\expect{0_M}{T_{\mu\nu}}{0_M} +  g^2 m^2  \expect{0_M}{\phi_m^\dagger
T_{\mu\nu} \phi_m}{0_M}\ -
\nonumber\\ &&2 g^2m^2{\rm Re} \left ( \expect{0_M}
{ T_{\mu\nu} \int \! d\tau_2 \int^{\tau_2} \! d \tau_1 e^{-i m \tau_2}
f(\tau_2) \phi(\tau_2) e^{+i m \tau_1} f^*(\tau_1) \phi(\tau_1) }{0_M}
\right)\nonumber\\ \label{threefivexi} \end{eqnarray}

It is  convenient to reexpress the last term in eq. \ref{threefiveix} and
eq. \ref{threefivexi} as  \begin{eqnarray}  &  2{\rm Re} \left [ \expect{0_M} {
T_{\mu\nu} \int \! d\tau_2 \int^{\tau_2} \! d \tau_1 e^{-i m \tau_2} f(\tau_2)
\phi(\tau_2) e^{+i m \tau_1} f(\tau_1)^* \phi(\tau_1) }{0_M} \right]
\nonumber\\ &\!\!\!\!\!\!\!\!\!\!\!\! =  {\rm Re} \left [ \expect{0_M} {
\phi_m^\dagger \phi_m T_{\mu\nu}
 }{0_M}\right] \nonumber\\ & +
 {\rm Re} \left [ \expect{0_M} { \left [ T_{\mu\nu} , \int \! d\tau_2
\int \! d \tau_1 \epsilon(\tau_2 - \tau_1) e^{-i m \tau_2} f(\tau_2)
\phi(\tau_2) e^{+i m \tau_1} f^*(\tau_1) \phi(\tau_1) \right]_- }{0_M}
\right] \nonumber\\ \label{threefivexii} \end{eqnarray}
where $\epsilon(\tau)= \theta(\tau) - \theta(-\tau)$.
The first term is
equal to the real part of eq. \ref{threefivevi}. This
stems from the fact
that
in both cases one is probing, through an
interference effect, the structure of the vacuum fluctuations that can excite
or deexcite the
atom. The second term takes into account that if the
atom was excited and then deexcited it necessarily occurred in that order.
Being a commutator, it
carries neither Minkowski energy nor Rindler energy
because
 Minkowski vacuum is an eigenstate of $H_M$
(eq. \ref{threethreev}) and $H_R$ (eq. \ref{threefouriv}). Furthermore it
vanishes when $(t_0,x_0)$ belongs to the intersection of $I_-$ and
$I_+$.
In view of these properties the only effect of this term is
 to redistribute
the flux density within the regions in causal contact.

\subsection{The uniformly accelerated two level atom}

We now evaluate the matrix elements
that appear in section 4.1
when the two level atom is uniformly
accelerated with acceleration $a$:
\begin{equation} t_a (\tau) =a^{-1} {\rm
sinh} a \tau \ ,\  x_a (\tau) =a^{-1} {\rm cosh} a \tau
\label{threefivexiibis}
\end{equation}

Consider first that $f(\tau)$ is equal to $1$ between $\tau_i$ and $\tau_f$
and that $\tau_f - \tau_i = T \rightarrow +\infty$ while $g^2 m T$ remains
finite. Then a direct golden rule calculation shows that the probability for
the uniformly accelerated atom to get excited is \cite{Unruh}
\begin{equation} P_{e,v} =  {1 \over 2} g^2 m
T N_m \label{threefivexiii}
\end{equation}
where $N_m = 1/ (e^{2 \pi m/a} -1)$ is the Bose Einstein distribution.

In this
limit
the operator $\phi_m^\dagger  \phi_m
$ appearing in eq. \ref{threefivevi} becomes the counting operator for
rindlerons of energy $m$: $a_{m,R}^\dagger  a_{m,R}$.
By getting excited, the  accelerated  atom
has post-selected that part of the vacuum wave
function $\ket{0_M}$ which contains a rindleron of
energy $m$.
This counting operator respects the bosonic statistics
of the field $\phi$ and differs from the
projector eq. \ref{threefourvii}
when acting
on states with two or more rindlerons of energy $m$. (It is only
when
 $m>>a$ , i.e.
in the Maxwell-Boltzman limit, that
these two operators and
their weak values eq. \ref{threefourviii} and
eq. \ref{threefivevi} coincide.
 This was precisely the case
studied in \cite{bmpps})

We turn now to the problem of a two level atom that interacts only during a
finite time with the field. First of all
we
notice that all the above matrix
elements of $T_{VV}(t_0,x_0)$ when  $(t_0,x_0)$ belongs to $I_-$ or
$I_+$ (except the second term of eq. \ref{threefivexii}) can be expressed
in terms of $P_{e,v}$ and of the following two functions  \begin{eqnarray}
\cc_+ (V) &=& \expect{0_M} {\phi(V) \phi_m^\dagger }{0_M} \nonumber\\ &=&
\int \! d\tau G_+(V,V_a (\tau)) e^{-i m \tau} f(\tau) \nonumber\\ \cc_-(V) &=&
\expect{0_M} {\phi(V) \phi_m}{0_M} \nonumber\\ &=& \int \! d\tau
G_+(V,V_a (\tau)) e^{+i m \tau} f^*(\tau) \label{threefivexiv} \end{eqnarray}
For instance,
discarding
 $ \expect{0_M}{T_{\mu\nu}}{0_M}$, eq. \ref{threefivevi}
and eq. \ref{threefivevii} read,
\begin{eqnarray} \bk{T_{VV}(I_-)}_{weak\  e} &=& {g^2 m^2 \over {
P_{e,v}}} \left( \partial_V \cc_+^* \right) \left( \partial_V \cc_-^*
\right)\nonumber\\ \bk{T_{VV}(I_+)}_{weak\  e} &=& {g^2 m^2 \over  P_{e,v}}
\left( \partial_V \cc_- \right) \left( \partial_V \cc_-^* \right)
\label{threefivexv} \end{eqnarray}
For $T_{VV}$ not to be singular the functions $ \partial_V \cc_+(V) $ and $
\partial_V \cc_-(V) $ must be regular (The last term in
eq. \ref{threefivexii} necessarily vanishes on the horizon $V=0$).
$ \partial_V
\cc_+(V) $ is equal to \begin{equation} \partial_V \cc_+(V) = -{1 \over 4
\pi} \int \! d \tau {1 \over V - a^{-1}e^{a\tau} - i \e} f(\tau)e^{-im\tau}
\label{threefivexvi} \end{equation}  It can be singular only for $V=0$ where
it takes the form  \begin{equation} -{1 \over 4 \pi} \int \! d \tau {1 \over -
a^{-1}e^{a\tau} - i \e } f(\tau)e^{-im\tau} \simeq  {a \over 4 \pi} \int \! d
\tau e^{-a\tau} f(\tau)e^{-im\tau} \label{threefivexvii} \end{equation} The
last integral is finite if and only if $f(\tau)$ decreases for $\tau
\rightarrow -\infty$ quicker than $e^{a\tau}$. Similarly if we had considered
right movers, the condition for finiteness on the future horizon would have
been sufficient rapid decrease of $f$ for $\tau \rightarrow +\infty$. Putting
all together the condition to not have singularities on the horizons is that

\begin{equation} \int \! d \tau {d t \over d \tau } \ \vert f(\tau)\vert =
\int \! dt \ \vert f(\tau(t))\vert < \infty \label{threefivexviii}
\end{equation}
The interaction of the atom with the field must last a finite
Minkowski time. When this is satisfied $f(\tau)$ decreases faster than
$e^{-a \vert \tau \vert}$ which implies that we are not considering a local
Lee model.

 \subsection{The weak values}

In order to obtain explicit expressions for $\cc_\pm$ we carry out the
following construction. First define the fourier transform of
$f(\tau)e^{-im\tau}$ by \begin{equation} f(\tau)e^{-im\tau} =
\int_{-\infty}^{+\infty}\!\! d\la { c_\la \over 2 \pi} e^{-i \la \tau}
\label{threefivexix} \end{equation} with the normalisation \begin{equation}
\int d \tau \vert f(\tau) \vert ^2 =  \int d\la { \vert c_\la \vert^2 \over 2
\pi} = T =\hbox{total time of interaction} \label{threefivexx} \end{equation}

The field operator $\phi_m$, the probability $P_e$,  the functions  $\cc_\pm$
and the expectation values of $T_{VV}$ can then be expressed in terms of
$c_\la$:
\begin{eqnarray} \phi_m &=& \int_0^\infty\!\! d \la  {a_{\la,R} \over \sqrt{4
\pi \vert \la \vert }}c_\la^*  \ + \  \int_{-\infty}^0\!\! d \la
{a_{\vert\la\vert,R}^\dagger  \over \sqrt{4 \pi \vert \la \vert }}c_\la^*
\nonumber\\ &=&\int_{-\infty}^{+\infty}\!\! d\la c_\la^* {1 \over \sqrt{4 \pi
\la (e^{\pi \la/a} - e^{-\pi \la /a})}}  (e^{\pi \la/2a} a_{\la,M} + e^{-\pi
\la/2a} a_{-\la,M}^\dagger ) \nonumber\\
\cc_+(V) &=&
\int_{-\infty}^{+\infty}\!\! d\la c_\la {1 \over \sqrt{4 \pi  \la (e^{\pi
\la/a} - e^{-\pi \la /a})}}
 e^{\pi \la/2a} \varphi_{\la,M}(V) \nonumber\\  &=&
\int_{-\infty}^{+\infty}\!\! d\la c_\la
{1 \over 4 \pi  \la}  \left[
( \tilde n_\la + 1) (aV)^{-i\la/a} \theta(V) +
\tilde n_\la e^{\pi \la/a}
\vert aV \vert^{- i\la/a} \theta(-V)
\right]\nonumber\\
 \cc_-(V)  &=& \int_{-\infty}^{+\infty}\!\! d\la c_\la^*
{1 \over 4 \pi \la}
\left[ \tilde n_\la  (aV)^{i\la/a} \theta(V) +
\tilde n_\la e^{\pi \la/a} \vert aV \vert^{i\la/a} \theta(-V)
\right] \label{threefivexxiii} \end{eqnarray}
Where
$\tilde n_\la = 1/ (e^{2 \pi \la/a}-1)$
is equal to (see eq. \ref{threethreexi})
\begin{eqnarray} \tilde n_\la &=&N_{\la}= \beta_{\la}^2
\quad \hbox{ for $\la > 0$}\nonumber\\
\tilde n_\la &=&-(N_{\vert\la\vert}+1)= -\alpha_{\vert\la\vert}^2
\quad \hbox{ for $\la
< 0$.} \end{eqnarray}

The contribution to the probability $P_{e,v}$ from left movers
reads
\begin{equation} P_{e,v} = g^2 m^2
\int_{-\infty}^{+\infty}\!\! d\la { \vert c_\la \vert^2 \over 4 \pi \la}
\tilde n_\la  \label{threefivexxiv} \end{equation}

As one picture is worth a thousand words we shall take a particular form for
$c_\la$ such that all the expressions can be evaluated explicitly.
\begin{equation}  c_\la = D { \la \over m} e^{-(\la - m )^2 T^2 /2}
 (1 -
e^{-2\pi \la /a})
 \label{threefivexxv} \end {equation} where $D$ is a
normalisation constant taken such as to verify eq. \ref{threefivexx}.
Then $f(\tau)$ reads
 \begin{eqnarray}
f(\tau)  &=&{D \over \sqrt{2 \pi}} {1 \over T} e^{-\tau^2/2T^2} \Bigl [ (1 - i
{\tau \over m T^2} ) - \nonumber\\ &&\quad e^{-2 \pi m/a}e^{2 \pi^2/a^2 T^2}
e^{i 2 \pi \tau / a T^2} (1 -  {i\tau + 2 \pi /a \over m T^2} ) \Bigr ]
\label{threefivexxvii}\end{eqnarray}
It clearly satisfies eq. \ref{threefivexviii}.

In order that the behaviour of the uniformly accelerated two level atom with
the coupling $f(\tau)$ given by eq. \ref{threefivexxvii} be physically
unambiguous, it is necessary that $c_\la$ be peaked around
$+m$
(the two level atom should be approximately a Lee model) and
the golden rule probability of transition
eq. \ref{threefivexiii} be recovered. If this is to
be the case then $T$ must satisfy
 $T >>  m^{-1} $ and $T>>a^{-1} $.
 The first condition is the usual demand that a wave packet be spread
over a distance at least equal to its inverse frequency. The second
condition, which corresponds to $T$ being greater than the euclidean
tunneling time $2 \pi a^{-1}$ \cite{pbt}, is required for the probability
$P_{e,v}$ to be linear in time and proportional to the Bose distribution
$N_m$. Then $ D \simeq
 2^{1/2}\pi^{1/4} T (N_m + 1)$ and
\begin{equation}
f(\tau) \simeq \pi^{-1/4}  e^{-\tau^2/2 T^2}
( 1+ N_m(1 - e^{i 2 \pi \tau / a T^2}))
\end{equation}
 We have
indicated throughout  by the symbol $=$ the exact
expressions for which no approximation has been made and by the symbol
$\simeq$ the approximate expressions valid when $T>> m^{-1}$
and $T>> a^{-1}$ as these last are
particularly easy to read and understand.

The weak values of $T_{vv}$ are readily obtained\footnote{Strictly
speaking, for the model eq. \ref{threefivexxvii}, $I_-$ does not include
regions where $V>0$ since the interaction lasts for an infinite Rindler
time. Nevertheless, since $f(\tau)$ decreases very quickly for $\tau \to
-\infty$, the
limiting value of $\bk {T_{vv}}_{weak\ }$ for $V>0$ as $u\to -\infty$
coincides with our expression.}
\begin{eqnarray}
\bk{T_{vv}(I_-, V > 0 )}_{weak\  e}
& =&
{ g^2 m^2 \over P_{e,v}}   \int\! d\la \!\int \! d\la^{\prime}\
 c_\la c_{\la^\prime}^{*}\
{1 \over (4 \pi)^2}
\ \tilde n_\la ( \tilde n_{\la^\prime} +1 )\
e^{-i(\la-\la^\prime)v} \nonumber\\
& =&
{m (N_m + 1)\over  2 \sqrt {\pi} T C_0}
 ( 1 - {i v + 2 \pi /a \over m T^2}) ( 1 + {i v \over m T^2})\
e^{ -{ (v-i \pi/a)^2 / T^2}  }
\nonumber\\ &\simeq & { m (N_m +1)\over 2 \sqrt{ \pi} T }
e^{ -{ (v-i \pi/a)^2/
T^2}  }
\label{weeki}\end{eqnarray}

\begin{eqnarray}
\bk{T_{vv}(I_-, V < 0 )}_{weak\  e}
&=&\bk{T_{vv}(I_+, V < 0 )}_{weak\  e}  \nonumber\\
&=& { g^2 m^2 \over P_{e,v}} \vert
\int d \la c_\la
{1 \over 4 \pi}
 { 1 \over e^{\pi \la /a} -
e^{- \pi \la /a}  } e^{-i\la v} \vert^2  \nonumber\\  &=& {m (N_m + 1)
\over  2 \sqrt {\pi} T C_0}
 \vert 1 - {i v +  \pi /a \over m T^2} \vert^2 e^{-{ v^2\over T^2}}
\nonumber\\ &\simeq& { m (N_m +1)\over 2 \sqrt{ \pi} T } e^{ - {v^2
\over T^2}}
\label{weekii}\end{eqnarray}

\begin{eqnarray}
 \bk{T_{vv}(I_+, V > 0 )}_{weak\  e}
&=&{ g^2 m^2 \over P_{e,v}} \vert \int d \la c_\la
{1 \over 4 \pi}
 \tilde n_\la e^{-i\la v} \vert^2   \nonumber\\
&=& {m N_m \over  2 \sqrt {\pi} T C_0}
 \vert 1 - {i v + 2 \pi /a \over m T^2}\vert^2   e^{-{ v^2\over
T^2}} e^{ 3\pi ^2 / a^2  T^2} \nonumber\\  &\simeq& { m N_m
\over 2 \sqrt{ \pi} T }
e^{-{ v^2\over T^2}}
 \label{weekiii}\end{eqnarray}
 where $C_0$ is a constant equal to
\begin{eqnarray}
 C_0 &=& (N_m + 1)^{-1}\left [ (1 - { \pi \over a m T^2}) - e^{- 2 \pi m/a}
e^{ 3\pi ^2 / a^2  T^2} (1 - { 2 \pi \over a m T^2})
\right] \nonumber\\ &\simeq&
1
\end{eqnarray}

We now present the
complementary Rindler and
Minkowski interpretations of the
 weak values of $T_{vv}$.

The Rindler description is that used by a uniformly accelerated observer in
the same quadrant as the two level atom. It is best understood by making
appeal to the isomorphism of the state of the field in the right Rindler
quadrant with an inertial thermal bath.

By getting excited the two level atom has selected that the thermal bath
contain at least one particle in the mode created by $\phi_m^\dagger$.
Furthermore energy flows for a massless field along the lines $u=cst$ and
$v=cst$. Therefore $\bk{T_{vv}(I_-, V>0 )}_{weak\  e}$ is centred around $v=0$
with at spread $\Delta v =T$ and carries a Rindler energy obtained by
integrating eq. \ref{weeki}
\begin{equation}
 \int dv \bk{T_{vv}(I_-, V>0 )}_{weak\  e} = {1 \over 2}{ \int d \la \vert
c_\la
\vert^2 \tilde n_\la ( \tilde n _\la +1 ) \over \int d \la \vert c_\la
\vert^2 {1 \over \la} \tilde n_\la }
\simeq {1 \over 2} m(N_m +1)
\label{seventytwo}
\end{equation}
The factor $N_m+1$ takes  correctly into account the Bose statistics of the
field since eq. \ref{seventytwo} corresponds to evalutating $<n^2>/<n>$
in a thermal
distribution. The factor $1/2$ arises since the atom could also have been
excited by
$u$ quanta.

Since by getting excited the two level atom has absorbed {\it one} quantum
the residual energy on $I_+$ is (see eq. \ref{weekiii})
\begin{equation}
 \int dv \bk{T_{vv}(I_+, V>0 )}_{weak\  e} = {1 \over 2}{ \int d \la
\vert c_\la
\vert^2 \tilde n_\la^2  \over \int d \la \vert c_\la
\vert^2 {1 \over \la} \tilde n_\la }
\simeq {1 \over 2} m N_m
\end{equation}

We now consider what is seen by a uniformly accelerated observer in the
left Rindler quadrant. Since Minkowski vacuum is an eigenstate of the total
Rindler energy (the boost operator) eq. \ref{threefouriv}, the Rindler
energy in the
left quadrant is equal to the energy in the right quadrant before the
transition occurs. Indeed integrating eq. \ref{weekii} and using the relation
$\tilde n_\la ( \tilde n_\la + 1) = 1/(e^{\pi \la /a} - e^{-\pi \la /a})^2$
yields
\begin{equation}
 \int dv \bk{T_{vv}(I_-, V>0 )}_{weak\  e} =
\int dv \bk{T_{vv}(I_-, V<0 )}_{weak\  e}
\end{equation}
The symmetry between the left and the right Rindler quadrants results in\\
$\bk{T_{vv}(I_-, V<0 )}_{weak\  e}$ being also centred around $v=0$ with width
$\Delta v=T$.
We have plotted these weak values of $T_{vv}$ in figure 1.

The Minkowski description, i.e. that used by an inertial observer, is best
understood by rewriting the weak value of $T_{VV}$ as
\begin{eqnarray}`
 \bk{T_{VV}(I_-)}_{weak\  e}
&=& {1 \over a^2 V^2}{ g^2 m^2 \over P_{e,v}} \int\! d\la \!\int
\!d\la^\prime c_\la^* c_{\la^\prime}
{1 \over 4 \pi} \sqrt{  \la \la^\prime
\tilde n_{\la^\prime} ( \tilde n_\la +1)
}
 \varphi^*_{\la,M}
\varphi^*_{-\la^\prime,M}
\nonumber\\ &=&
{1 \over a^2 V^2} {m (N_m + 1)\over  2
\sqrt{\pi} T C_0}( 1 + {i \over m a T^2} \ln (-aV-i\e) -{ \pi \over m a T^2})
\nonumber\\  &\ &
 \times  ( 1 - {i \over m a T^2} \ln (-aV-i\e) -{ \pi
\over m a T^2}) e^{-\left[ \ln (-aV-i\epsilon) \right]^2 / a^2 T^2}
\label{weekiv}
\end{eqnarray}
which we have sketched in figure 2. The $i \e$ is introduced only to define
$\ln (-aV-i\e)$ as $\ln \vert aV \vert $ for $V<0$ and as
$\ln \vert aV \vert
- i \pi$ for $V>0$. Upon taking the limit $\e
\rightarrow 0$ no singularity occurs.
In fact
$\bk{T_{VV}(I_-)}_{weak\  e}$ given in eq. \ref{weekiv} vanishes for $V=0$.
This is an accident due to the particular form of $c_\la$ chosen in
eq. \ref{threefivexxv} (it has
zero's for $\la=i n a$, $n= ..,-1,0,1,..$).
But, from the expression for $\cc_{\pm} (V=0)$ given
in eq. \ref{threefivexvii}, it
results that the generic behaviour of $T_{VV}$ is to stay finite as $V
\to 0$. In more physical terms this corresponds to saying
that
the Minkowski vacuum fluctuation that induces the transition straddles
the
horizon with no clear cut separation between the pieces in the left and
right
quadrants.

In $I_-$, the total Minkowski energy (the integral of
eq. \ref{weekiv} with respect to $V$) vanishes since $\ket{0_M}$ is
an eigenstate of
$H_M$ (eq. \ref{threethreev}). In other words the total
Minkowski energy does
not fluctuate and is always equal to its eigenvalue zero: vacuum
fluctuations carry no energy.
The Minkowski energy in the region $V<0$ is real and positive therefore the
energy in the region $V>0$ must integrate to an exactly compensating real
and negative value. This is not
in contradiction with the positivity of Rindler energy in the right quadrant
since the expressions for the Rindler and the Minkowski energy differ by the
jacobian ${ dv / dV} = {1 / aV}$. The oscillations of $T_{vv}$ for
$V>0$ that occur in eq. \ref{weeki} as $v \rightarrow - \infty$ (which are
negligible in
the Rindler description) are dramatically enhanced by the jacobian in such a
way that the Minkowski energy in the right quadrant becomes negative.

In $I_+$, after the atom has made a transition, the Minkowski energy takes the
form \begin{eqnarray}
\bk{T_{VV}(I_+)}_{weak\  e} &=&
{1 \over a^2 V^2}{ g^2 m^2 \over
P_{e,v}} \vert \int\! d\la c_\la
\sqrt{ {
\la \tilde n_\la \over 4 \pi}} \varphi_{-\la, M}^*
\vert^2\nonumber\\
&=&{1 \over a^2 V^2} {m (N_m + 1)\over  2 \sqrt{\pi} T C_0} \vert 1 -
{i \over m a
T^2} \ln (-aV-i\e) -  {\pi\over m a T^2}  \vert^2
\nonumber\\
&&\quad\quad
  \vert e^{-{\left[ \ln (-aV-i\epsilon) \right]^2 / a^2 T^2}} \vert^2 \vert
e^{-i{m }\ln (-aV-i\epsilon)/ a } \vert^2  \end{eqnarray}

It is manifestly real and positive. This is as it should be since we are
calculating the mean value of the energy in a state that is not Minkowski
vacuum.
By absorbing the positive Rindler energy $m$, the two level atom has reduced
the negative Minkowski energy on the right.

These results will be used when analysing the mean fluxes emitted by an
accelerated oscillator. This is the subject of the next chapter.

\section{Fluxes and Particles Emitted by an Accelerated Oscillator}

\subsection{introduction}

We analyse the mean fluxes emitted during the
thermalisation period (i.e. when the initial state is the ground state)
and in
thermal equilibrium. The analysis is first carried out to order $g^2$
using the
model of chapter 4. In a
second stage we  use the model introduced by Raine, Sciama and Grove (RSG)
\cite{RSG}
to prove that the
various properties characterising thermal equilibrium previously obtained
in $g^2$ are recovered to all orders in $g$.

We briefly sketch the main results.
During thermalisation a
steady flux of negative Rindler energy is emitted
 ( $\bk{ T_{vv}} \simeq - g^2 m^2
N_m /2$). This is understood from the isomorphism \cite{Grove}
 with the thermal bath: as the
atom gets exited it absorbs energy from the thermal bath, thus the minus sign.
The transcription of this flux to Minkowski quanta is more subtle. Oscillatory
tails in the Rindler flux are enhanced by the jacobian that converts from
Rindler to Minkowski energy with the net result that positive Minkowski energy
is emitted.
In the Minkowski description the origin of
the steady negative flux is due to a ''repolarisation"
of the atom corresponding to
the fact that the probability of finding the atom in its exited level
decreases with time. This repolarisation is similar (CPT conjugate)
 with that which occurs
when negative energy is absorbed by an inertial detector \cite{Grove2}.

When thermal equilibrium is reached the emission
of Rindler energy ceases
because the aborption of Rindler energy associated with exciting the atom is
exactly compensated (except for oscillatory transients) by the emission
provoked by
the inverse quantum jump.
Nevertheless in this equilibrium situation there is
a net production of Minkowski energy because both absorption and
emission of Rindler quanta correspond to emission of Minkowski quanta.
These Minkowski quanta interfere in such a way that their energy content is
located only at the ends of the interaction period (in the oscillatory
transients).
The compatibility of the two descriptions arises, once more, from
 the time dependence of
the Doppler shift relating Rindler and Minkowski frequencies. It
is remarkable
that this Doppler factor ${dV /dv} = e^{av}$ both
leads
to the thermalisation of the accelerated atom
through the Bogoljubov transformation eq. \ref{threethreex} and
allows the conciliation of the two descriptions of the emitted
flux.

\subsection{Fluxes and particles in $g^2$ during thermalisation}

The analysis is first carried out for the adiabatic switch on and off used in
chapter 4 (to reveal the oscillatory tails)
 and then for a sudden switch on and off (to display the stationary
regime).

The mean energy radiated by the uniformly accelerated two level atom
is given by (see eq. \ref{threefivex} et seq.)
\begin{eqnarray}
 \bk{T_{vv}}_e &=& P_e \bk{T_{vv}(I_-)}_{weak\  e} +
P_g \bk{T_{vv}(I_-)}_{weak\  g}
\nonumber\\  &=&  g^2 m^2  \expect{0_M}{\phi_m^\dagger
T_{vv} \phi_m}{0_M}\ -  g^2m^2
{\rm Re}  \left ( \expect{0_M} {
\phi_m^\dagger \phi_m T_{vv}
 }{0_M}\right) \nonumber\\
&&-g^2m^2\ \times\ \hbox{last term of eq. \ref{threefivexii}}
\label{rayi}
\end{eqnarray}
The third term carries neither Rindler nor
Minkowski energy, it
is non vanishing only in the causal future of the atom and will
be discarded.
The first two terms taken separately are non vanishing outside
of the causal future of the two level atom but upon taking their sum
causality
is restored. Their sum reads
\begin{eqnarray}
\bk{T_{vv}(I_+)}_e  &=& -
g^2m^2\int d\la \int d\la^\prime c_\la c^*_{\la^\prime}
{1 \over (4 \pi)^2} (\tilde n_\la + \tilde n_{\la^\prime})
e^{-i(\la-\la^\prime)v}
\nonumber\\
&\simeq&{-
{g^2 m^2} \over 2  }
N_m  { e^{-v^2/T^2} \over\pi^{1/2}}
[ (N_m + 1) \cos (2 \pi v/aT^2) - N_m]
\label{rayii}
\end{eqnarray}
As announced it carries negative
Rindler energy:
 \begin{eqnarray}
\int\! dv \bk{T_{vv}(I_+)}_e &=& -  {g^2
m^2 \over 4 \pi} \int
d\la
\vert c_\la \vert^2  \tilde n_\la
 \nonumber\\  &\simeq&
-{1 \over 2}g^2 m^2 N_mT= -m P_{e,v}
\label{rayiii}
\end{eqnarray}

The total Minkowski energy radiated is
computed by integrating eq. \ref{rayii}:
\begin{eqnarray} \int
dV \bk{T_{VV}(I_+)}_e &=&
 \int dv  e^{av} \bk{T_{vv}(I_+)}_{ e}\nonumber\\ &\simeq& {1 \over
2} g^2 m^2   N_m T e^{a \tau_0}(1 + 2 N_m)= +m P_{e,v} e^{a \tau_0}
(1 + 2 N_m)
\label{rayiv}
\end{eqnarray}
where $ e^{a \tau_0}$ is the mean Doppler
effect associated with the window
function $f(\tau)$ eq. \ref{threefivexxvii}:
 \begin{equation}
\int\! dv e^{-av} e^{-v^2/T^2} \cos(2 \pi v/aT^2) \simeq - e^{a \tau_0}
\ .
\label{rayv}\end{equation}
Note that the sign flip in eq. \ref{rayv} of the Minkowski energy versus the
Rindler energy can be conceived as arising from the imaginary part of
the saddle point
of eq. \ref{rayv}: $v_{sp} = -a T^2/4 + i \pi /a$ and is for that reason very
similar to the flip of frequency which leads through a tunnelling amplitude
to a non vanishing $\beta$ coefficient (see \cite{pbt}). The additional
factor $2N_m + 1$ comes probably from the particular switch off function
$f(\tau)$.

In order to get a more precise picture of the physics involved we
now analyse the case where the time dependent coupling is
$f(\tau) = \theta(\tau) \theta (T - \tau)$. With this time dependence the
transients are singular (as is seen by considering the Fourier transform of
the $\theta$ function) and will not be studied. On the contrary the steady
part is easily computed and corresponds exactly to the intermediate values
($- aT^2  << \tau << a T^2   $)
found in the adiabatic situation described
above in eq. \ref{rayii}. In addition it shows
explicitly the  relations that exist between the
flux emitted and the  transition rate (not only the probability as in
eq. \ref{rayiii}).

We first compute, using standard perturbation theory, the relevant
formulae for the probability of transition, for
the rate of transition and for the emitted flux. The probability of
spontaneous emission (due to the $v$-modes only) is given by
 \begin{eqnarray}
P_{e,v}(T) &=& g^2 m^2 \int_0^T d\tau_1 \int_0^T d\tau_2
e^{-im(\tau_2 - \tau_1)}
\bk{\phi(\tau_2)\phi(\tau_1)} \nonumber\\ &\simeq& {1 \over 2} g^2 m  N_m
T
\label{rayvi}
\end{eqnarray}
The second line contains the golden rule result valid when
$aT\to \infty$ with $g^2 T$ finite. It is useful for the following to
introduce the rate of transition, the derivative of $P_{e,v}(T)$:
\begin{eqnarray}
\dot P_{e,v}(T) &=& {d P_{e,v}(T) \over dT}
= g^2 m^2 2 {\rm Re} \left[
\int_0^T d\tau e^{-im(T-\tau)} \bk{\phi(T)\phi(\tau)}\right] \nonumber\\
&\simeq& {1 \over 2} g^2 m N_m
\label{rayvii}
\end{eqnarray}
On the left hand side of the
accelerated trajectory, this rate is related to the (steady part of) the
stress energy tensor: \begin{eqnarray} &\ &\quad\quad\bk{ T_{vv}(v
=T)}_e =\nonumber\\ &=& g^2 m^2 2 {\rm Re} \left[ \int_0^T d\tau_2
\int_0^{\tau_2} d\tau_1 e^{-im(\tau_2 - \tau_1)} \bk{ [\phi(\tau_2) ,
T_{vv}(T) ] \phi(\tau_1)}\right] \nonumber\\ &=& g^2 m^2 2 {\rm Re} \left[
\int_0^T d\tau e^{-im(T-\tau)} \bk{i \partial_v \phi(T) \phi(\tau)}\right]
\nonumber\\ &=& - m \dot P_{e,v}(T) + g^2m^2 2 {\rm Re} \left[ i e^{-imT} \bk{
\phi(T) \phi(0)}\right]
\label{rayviii}\end{eqnarray}
 The first equality follows
straightforewardly from the expansion of the evolution operator $e^{-i\!
\int\! H_{int} d \tau}$ in $g^2$. The second equality is obtained using
$[\phi(\tau_2) , T_{vv}(\tau_1) ] = i \partial_v \phi \delta (\tau_1 -
\tau_2)$. The third equality follows by using

 $\bk{ \partial_{\tau_1}
\phi(\tau_1) \phi(\tau_2)} =- \bk{  \phi(\tau_1)
\partial_{\tau_2}\phi(\tau_2)}$  and integrating by parts. The final result
contains a steady part proportional to $- m \dot P_{e,v}(T)$ which
 tends  to $-{1
\over 2} g^2 m^2 N_m$ in the golden rule limit and an oscillatory term
(which is exponentially damped if a slight mass is given to $ \phi$). The
steady piece simply indicates that to an increase of the probability
 to make a
transition corresponds the absorption of the necessary Rindler energy to
provoke this transition.

These expressions are now decomposed in terms of the Minkowski
basis $e^{-i\omega V}/\sqrt{4 \pi \omega}$. The probability of
transition eq. \ref{rayvi}
reads
 \begin{eqnarray} P_{e,v}(T) &=&
g^2 m^2 \int_0^\infty\! d\omega \ \vert\! \int_0^T\! d\tau\ e^{-im\tau} {
e^{-i{\omega \over a} e^{a \tau}} \over \sqrt{4 \pi \omega}} \vert^2
\nonumber\\ &=& \int_0^\infty\! d \omega\ P_{e,v,\omega}(T)
\label{rayix}
\end{eqnarray}
Similarly the transition rate eq. \ref{rayvii} becomes
\begin{eqnarray} \dot P_{e,v}(T)
 &=& g^2
m^2 \int_0^\infty\! d\omega \ 2 {\rm Re} \left[ \int_0^T d\tau e^{-im(T-\tau)}
{ e^{-i{\omega \over a} (e^{aT} - e^{a \tau})} \over 4 \pi \omega} \right]
\nonumber\\ &=& \int_0^\infty \! d\omega\  \dot P_{e,v,\omega}(T)
\label{rayx}
\end{eqnarray}
And the total Minkowski energy is given by
 \begin{eqnarray} H_e(T) &=&
\int_{-\infty}^{+\infty} dv e^{av} \bk{T_{vv}}_e
 \nonumber\\ &=& \int_0^\infty\! d\omega\ \omega P_{e,v, \omega}(T)
\label{rayxi}
\end{eqnarray}
Where in the first equality the integral is only over region of positive $V$
since by causality the mean energy is unaffected in the other quadrant. The
second equality follows from the diagonal character of the energy operator.
The positivity of
$H_e(T)$
is manifest since
all the $P_{e,v,\omega}(T)$
are
positive definite. Nevertheless the time derivative of $H_e(T)$
is negative, within the steady regime,
\begin{eqnarray}
{dH_e \over dT} &=& \int_0^\infty\! d\omega\ \omega
\dot P_{e,v,\omega}(T)\nonumber\\ &=& e^{a v(T)} \bk{T_{vv}(v(T))}\nonumber\\
&=& -m e^{av(T)}  \left[ \dot P_{e,v} (T) + \hbox{oscillatory "damped" term}
 \right]
\label{rayxii}
\end{eqnarray}
${dH / dT}$
negative implies that,
for large $ \omega$ (since $ \dot P_e(T) >0$), some $\dot P_{e,\omega}$
are negative. This
corresponds to a ''repolarisation" since all the $P_{e,\omega}$
are positive definite
and vanish for $\tau \leq 0$. This repolarisation is
exactly the inverse process of the absorption of negative energy by an atom
\cite{Grove2}.

\subsection{Fluxes and particles in $g^2$ at equilibrium}

Before studying the equilibrium situation it behoves us first to consider the
flux emitted by an atom that makes a transition from excited to ground state.

The probability (due to the $v$-modes) that a uniformly accelerated
two level atom initially in its
exited state ends up in its ground state is
\begin{eqnarray}
 P_{d,v} &=& g^2 m^2
\expect{0_M} {\phi_m \phi_m^\dagger}{0_M}
\nonumber \\ &\simeq& {1 \over 2}g^2 m
 (N_m
+1) T
\label{rayxiii}
\end{eqnarray}
The mean energy emitted is, in the adiabatic switch off case,
\begin{eqnarray}
\bk{T_{vv}}_d &=&   g^2 m^2
\expect{0_M}{\phi_m  T_{vv} \phi_m^\dagger}{0_M}\  -
g^2m^2{\rm  Re} \left [ \expect{0_M} {\phi_m \phi_m^\dagger T_{vv}}{0_M}
\right] \nonumber\\  && -g^2m^2\ \times\
\hbox{last term of eq. \ref{threefivexii}}
\label{rayxiv}
\end{eqnarray}
The sum of the first two terms reads
\begin{eqnarray}
\bk{T_{vv}}_d &=&
 g^2 m^2
\int\!
d\la \int \! d\la^\prime c_\la c_{\la^\prime}^*
{1 \over (4 \pi)^2}
\left (  \tilde n_\la + \tilde n_{\la^\prime}  + 2 \right)
e^{-i(\la-\la^\prime)v}\nonumber\\
&\simeq&{g^2 m^2 \over 2 \sqrt{\pi}}
(N_m + 1) e^{-v^2/T^2}
[ 1 - N_m \{ \cos (2 \pi v/a T^2) -1 \} ]
\label{rayxv}
\end{eqnarray}
and the total Rindler energy radiated is
\begin{eqnarray}
\int\! dv \bk{T_{vv}(I_+)}_d &=&   {g^2
m^2 \over 4 \pi} \int
d\la
\vert c_\la \vert^2  (\tilde n_\la + 1)
\nonumber\\  &\simeq&
{1 \over 2} g^2 m^2 (N_m+1)T   = m P_{d,v}
\label{rayxvi}
\end{eqnarray}
In the example for which the time dependent
coupling is $f(\tau) = \theta(\tau) \theta( T- \tau)$, one finds
the following relation between the derivative of the probability
$ \dot P_{d,v}(T)$ and the flux $ \bk{T_{vv}}_d$:
 \begin{equation}
 \bk{T_{vv}(T)}_d =
+ m \dot P_{d,v}(T) + \hbox{oscillatory ''damped" term}
\label{rayxvii}
\end{equation}
The sign in front of $ \dot P_d(T)$ is now positive (contrary to the one
in eq. \ref{rayviii}).
Deexcitation  consists in emitting the energy stored in the atom.

The total
Minkowski
energy emitted
is obtained by integrating the first term of eq. \ref{rayxiv} only:
\begin{eqnarray}
 \int dV \bk{T_{V V}}_d  &=& \int dV g^2 m^2
\expect{0_M}{\phi_m  T_{VV} \phi_m^\dagger}{0_M}
 \nonumber\\ &\simeq&
{g^2 m^2 \over 2}(N_m+1)T e^{a \tau_0}(2 N_m + 1) =
m P_{d,v} e^{a \tau_0}(2 N_m + 1)
\end{eqnarray}
For the deexcitation, the integrated Rindler and Minkowski energies
have the same sign and are related by the mean Doppler shift
$e^{a \tau_0}$ times $(2 N_m + 1)$.

We now turn to the thermal equilibrium situation. The occupation
probabilities, $p_0$ for the ground state and $p_1$ for the excited one,
are related to the transition rates by the Einstein relations:
\begin{equation} {p_1 \over p_0} = {\dot P_e \over \dot P_d} =
{N_m \over N_m+1} = e^{-2 \pi m/a}
\label{rayxix}\end{equation}

The energy radiated is the sum of the fluxes
emitted when the atom is initially
in its ground state
and
when the atom is initially in its
exited state weighted by
their initial probabilities. This stems from the fact that
the energy momentum operator changes the photon number by an even number and
that the interaction hamiltonian changes the photon number by an odd number
while changing the state of the atom.
Hence one has
 \begin{eqnarray}
\bk{T_{vv}}_{equil} &=&
p_0  \bk{T_{vv}}_e +  p_1  \bk{T_{vv}}_d
 \nonumber\\ &\simeq&
-m p_0  \dot P_{e,v} + m p_1  \dot P_{d,v} =  0
\label{rayxx}
\end{eqnarray}
The steady fluxes cancel exactly because of thermal
equilibrium.
This is Grove theorem in $g^2$ \cite{Grove} \cite{mpbrsg}.
 Only the oscillatory transients remain.
They read for the smooth switch on and off
\begin{eqnarray}
 \bk{T_{vv}}_{equil} &=&  g^2 m^2
{1 \over 2 N_m +1}
  \int d\la \int d\la^\prime  c_\la
c_{\la^\prime}^*
{1 \over (4 \pi)^2}
\nonumber\\
&\ &\quad\quad
\left[
N_m ( \tilde n_\la +\tilde n_{\la^\prime} + 2 )
- ( N_m + 1)  (\tilde n_\la +\tilde n_{\la^\prime} )
\right ]
e^{-i(\la- \la^\prime) v}\label{rayxxi}
\end{eqnarray}
where $1/(2N_m+1)$ comes from the normalisation of probabilities:
$ p_0 +  p_1=1$.
 We have sketched $\bk{T_{VV}}_{equil}$ and
$\bk{T_{vv}}_{equil}$ in figure 3.
The total Rindler
energy emitted is then
\begin{equation}
 \int\! dv\ \bk{T_{vv}}_{equil} =
{ g^2 m^2
\over 4 \pi}{1 \over 2 N_m + 1}
 \int d\la \vert c_\la \vert^2 (N_m - \tilde n_\la )
\label{rayxxii}
\end{equation}
It tends to zero when $c_\la$ tends to a $\delta$ function
as the time of interaction
tends to $\infty$.
However, the total Minkowski energy
{\it increases}
with time and is
given by
 \begin{eqnarray} \int dV \bk{T_{VV}}_{equil} &=&
p_0  \int dV \bk{T_{VV}}_e +  p_1 \int dV \bk{T_{VV}}_d
\nonumber\\ &\simeq&
m( p_0 \dot P_{e,v} + p_1 \dot P_{d,v}) T e^{a \tau_0}(2 N_m + 1)
\label{rayxxiii}
\end{eqnarray}
The Minkowski energy of the two fluxes coincide and sum up.
Notice that this result equals the "naive" guess which is:
The total energy is the integral over the interacting period
of the rate of transition times the varying Doppler shift
times the energy gap $m$. We now go to all order in $g$ to prove that
this emission of Minkowski quanta is not an artefact of the second
order perturbation theory.

\subsection{Particles and Fluxes to all order in $g$}

  We use the exactly solvable model of RSG
\cite{RSG} \cite{mpbrsg} \cite{Unruh92}
to prove that one does recover, to all
order in $g$, that every quantum jump of the
accelerated oscillator, in thermal equilibrium in Minkowski vacuum,
leads to the emission of a Minkowski quantum. Hence the rate of
production of the Minkowski quanta is simply the rate of internal
transitions of the oscillator. But, as in second order perturbation
theory, these quanta interfere
and their
energy
content is found
(due to the complete neglection of the recoil)
 at the edges of the interacting period
only.
Then we give a general proof
that the
stationary thermal Rindler equilibrium corresponds to a
production of Minkowski quanta.

We first recall the main properties of the RSG model and then
analyse the particle content of
the emitted fluxes.

The action of this system consisting
of a massless field coupled to
a harmonic oscillator maintained in constant acceleration
is
\begin{eqnarray}
S &=&\int\! dt dx\ \left[ {1 \over 2} \left[ ( \partial_t \phi)^2
- ( \partial_x \phi)^2  \right] \right.
\nonumber\\
&&\quad + \left.
 \int\! d\tau\   \left[ { 1 \over 2} \left[ ( \partial_{ \tau} q)^2 - m ^2
q^2  \right] + e( \partial_{ \tau} q)  \phi  \right]
\delta ^2 (X^{\mu } -X^{\mu
}_a (\tau ))\right]
\label{un}
\end{eqnarray}
where $X^{\mu } (\tau )$ is the accelerated trajectory
eq. \ref{threefivexiibis} and $e= g \sqrt{2 m}$ is a rescaled coupling
constant. Since this action is quadratic, the Heisenberg equations are
identical to the classical Euler Lagrange ones. They read:
\begin{equation}
 \partial_u  \partial_v  \phi = {e \over 4 } \theta(V) \delta
( \rho - 1/a)  \partial_{ \tau} q
\label{deux}
\end{equation}
\begin{equation}
\partial_{ \tau} ^2 q +  m ^2
q = - e \partial_{ \tau}  \phi(X^{\mu } (\tau ))
\label{trois}
\end{equation}
The left part of the field (i.e. for $V<0$)
is, by causality, identically free.
And, for $V>0$, on the left of the
accelerated oscillator trajectory, the $v$-part of the field only is
scattered. There the general solution is
\begin{equation}
\tilde \phi (u,v) = \phi (u) + \phi (v) + {e\over 2 } \tilde q(v)
\label{quatre}
\end{equation}
\begin{equation}
\tilde q(v)= q(v) + i  \int _{-\infty}^{+\infty} \! d\la\
 \psi_{ \la} e^{-i \la v}
 \left[ \phi _{ \la,R,v} + \phi _{ \la,R,u} \right]
\label{cinq}
\end{equation}
where $ \phi (u)$ and $ \phi (v)$ are the
homogenous free solutions of  eq. \ref{deux}; where the
operator $ \phi _{ \la,R,v}$ is defined by
\begin{eqnarray}
 \phi _{ \la,R,v} &=&  \int\! {dv \over 2 \pi } e^{i \la v}  \phi (v)
\nonumber\\ &=& {1  \over \sqrt{4 \pi | \la |}}  \left[ \theta ( \la)
a_{\la,R} +  \theta (- \la) a_{- \la,R}^\dagger  \right]
\label{six}
\end{eqnarray}
(a similar equation defines $ \phi _{ \la,R,u}$);
where $ \psi _{ \la}$ is given by
\begin{equation}
\psi _{ \la} = { e\la \over  m ^2 -  \la ^2 -i e^2  \la /2}
\label{sept}
\end{equation}
and where $q(v)$ is a solution of
\begin{equation}
\partial_{ \tau} ^2 q +  m ^2 q + {e^2  \over 2} \partial_{
\tau}q = 0
\label{huit}
\end{equation}
The two independent solutions of  eq. \ref{huit} are exponentially
damped as $ \tau$
increases. Being interested by the properties at equilibrium, we drop
$q(v)$ from now on. Then, the remaining part of $ \tilde q(v)$ is a function
of the free field only. Hence, in Fourier transform, eq. \ref{quatre}
reads
\begin{eqnarray}
\tilde  \phi _{ \la,R,u} &=&  \phi _{ \la,R,u}
\nonumber\\
\tilde  \phi _{ \la,R,v} &=&  \phi _{ \la,R,v} (1+ i{e\over 2} \psi _{ \la})+
(i{e\over 2} \psi _{ \la})  \phi _{
\la,R,u}
\label{neuf}
\end{eqnarray}
The second term in eq. \ref{neuf} mixes $u$ and $v$ modes. It encodes the
static Rindler polarisation cloud (see \cite{Unruh92} \cite{mpbrsg})
which accompanies the
oscillator and carries
neither
Minkowski nor Rindler energy. In order to simplify the following
equations, we drop it and multiply the other scattered term by two
for unitary reason -see
below. (By a simple and tedious algebra, one can
explicitly verify that this modification does not
affect the main properties
of the emitted fluxes). Then  eq. \ref{neuf} becomes
\begin{equation}
\tilde  \phi _{ \la,R,v} =  \phi _{ \la,R,v} (1+ ie\psi _{ \la})
\label{dix}
\end{equation}
It is useful, for future discussions, to introduce explicitly the
scattered operators  $\tilde a_{\la,R}$,
and the scattered modes $\tilde \varphi_{\la,R} (v)$
\begin{equation}
\tilde a_{\la,R} =  < \varphi_{\la,R} |  \tilde  \phi > =
a_{\la,R}(1+ ie\psi _{ \la})
\label{onze}
\end{equation}
\begin{equation}
\tilde \varphi_{\la,R} (v) =- \left[  a_{\la,R}^\dagger,  \tilde  \phi
(v)
\right]_- = (1+ ie\psi _{ \la}) \varphi_{\la,R} (v)
\label{douze}
\end{equation}
whereupon the scattered field operator $\tilde \phi (v)$ may be written as
\begin{eqnarray}
\tilde \phi (v)  &=& \int_0^{\infty}\! d\la  \left[ \tilde a_{\la,R}
\varphi_{\la,R} + h.c.  \right]
\nonumber\\  &=& \int_0^{\infty}\! d\la
\left[ a_{\la,R} \tilde \varphi_{\la,R} +  h.c. \right]
\label{treize}
\end{eqnarray}
It is now straitforward to obtain the  scattered Green function and its
Rindler energy content. If the initial (Heisenberg) state is Minkowski vacuum
the $v$-part of the scattered Green function is, for $V,V^\prime > 0$,
\begin{eqnarray}
\tilde G_+(v,v^\prime) &=& \expect{0_M}{\tilde  \phi(v)
 \tilde \phi(v^\prime)}{0_M}\nonumber\\ &=&  \int_0^{\infty}\! d\la
 |1+ ie\psi _{ \la}|^2 \left(
\beta_{\la}^2
 \varphi_{\la,R}^*(v)  \varphi_{\la,R}(v^\prime) +
\alpha_{\la}^2
 \varphi_{\la,R}(v)  \varphi_{\la,R}^*(v^\prime) \right)
\nonumber\\
&=& G_+(v,v^\prime)
\label{quatorze}
\end{eqnarray}
where $G_+(v,v^\prime)$ is the unperturbed Minkowski Green function and
where we have availed ourselves of the identity (see eq. \ref{sept})
\begin{equation}
|1+ ie\psi _{ \la}|^2 = 1
\label{quinze}
\end{equation}
This unitary
relation
expresses the conservation of the number of Rindler
particles.
Indeed there is no mixing of positive and negative frequencies in
eq. \ref{onze}; in other words, the $ \beta$-term of the ''Bogoljubov"
transformation eq. \ref{onze}
vanishes.

The identity of the Green functions in eq. \ref{quatorze} proves that, once the
the steady regime is established,
no flux
is, {\it
 in the
mean},
emitted.
This is
Grove theorem \cite{Grove} \cite{RSG}.

We now examine how this stationary scattering of Rindler modes is
perceived in Minkowski terms.
 The Minkowski scattered modes $ \tilde \varphi_{\la,M}$ are given by
\begin{eqnarray}
 \tilde \varphi_{\la,M} &=&  - \left[ a_{\la,R}^\dagger,  \tilde \phi (V)
\right]_-
\nonumber\\ &=&
\varphi_{\la,M} (1+ie\alpha _{\la} ^2 \psi_{\la}) -ie\alpha _{\la}
\beta_{\la} \psi_{\la} \varphi_{- \la,M} ^*
\nonumber\\ &=&  \tilde  \alpha _{\la} \varphi_{\la,M} +  \tilde
\beta_{\la} \varphi_{- \la,M} ^*
\label{seize}
\end{eqnarray}
\begin{eqnarray}
\tilde \varphi_{- \la,M} &=& \varphi_{- \la,M} (1 -ie\beta_{\la}^2
\psi_{- \la}) -ie\alpha _{\la} \beta_{\la} \psi_{- \la} \varphi_{ \la,M}
^* \nonumber\\ &=&  \tilde  \alpha _{- \la} \varphi_{- \la,M} +  \tilde
\beta_{- \la} \varphi_{\la,M} ^*
\label{dsept}
\end{eqnarray}
where $ 0 < \la < \infty $ and where we have introduced the scattered
Bogoljubov coefficients:
\begin{eqnarray}
 \tilde  \alpha _{\la} &=&  1+ie\alpha _{\la} ^2 \psi_{\la}
\nonumber\\
 \tilde
\beta_{\la} &=& -ie\alpha _{\la}
\beta_{\la} \psi^*_{\la}
\nonumber\\
\tilde  \alpha _{- \la} &=& 1 +ie\beta_{\la}^2
\psi^*_{\la}
\nonumber\\
\tilde
\beta_{- \la} &=& -ie\alpha _{\la} \beta_{\la} \psi_{\la}
\label{dhuit}
\end{eqnarray}
One verifies that the unitary relation is satisfied:
$| \tilde \alpha_{\la}|^2-| \tilde \beta_{\la}|^2=1$.
 The fact that the $ \tilde \beta $ are different from zero indicates
that each couple of jumps of the oscillator
(the absorption and subsequent emission of a Rindler quantum)
leads, in Minkowski vacuum, to the production of two Minkowski quanta.
The member $ \varphi_{- \la , M}$ is emitted when the oscillator
absorbs a Rindleron and jumps into a higher level and the other one,
$ \varphi_{ \la , M}$ is emitted during the inverse process.
This is
manifest
in the mean energy flux:
\begin{eqnarray}
\tilde T_{VV} &=&  \lim_{V^\prime
\rightarrow V}\partial_V \partial_{V^\prime} \left[ \tilde \phi(V)
\tilde \phi(V^\prime) -
\phi(V) \phi(V^\prime) \right]
\nonumber\\ &=& 2  \int_{-\infty}^{+\infty} \! d\la\  | \tilde \beta_{\la}|^2
| \partial_V \varphi_{\la,M}|^2 +
Re \left[ \tilde  \alpha
_{\la} \tilde \beta_{\la}^*
\partial_V \varphi_{\la,M} \partial_V \varphi_{- \la,M} \right]
\label{dneuf}
\end{eqnarray}
where upon the total Minkowski energy, given by the integration of the
first term only, is:
\begin{eqnarray}
\tilde H_M &=&  \int_{-\infty}^{+\infty} \! dV
\tilde T_{VV}
\nonumber\\ &=&  \int_{0}^{+\infty}  d\la\ \la (|
 \tilde \beta_{\la}|^2 +| \tilde \beta_{- \la}|^2)
  \int_{-\infty}^{+\infty} \! {dV \over 2 \pi} {1 \over a^2 |V+i
\epsilon|^2 }
\label{vingt}
\end{eqnarray}
Exactly as in second order perturbation theory,
there is a steady regime during which all the emitted quanta interfere
destructively
leaving no contribution to the
{\it
mean
}
flux (see eq. \ref{quatorze}). But all non diagonal matrix
elements will be sensitive to the
created pairs. This is also the case for the
the total energy eq. \ref{vingt} since being diagonal in
$ \omega$
it ignores the destructive interferences (the second term of eq. \ref{dneuf}
whose role is to make the mean flux vanishing during the steady regime).

In order to prove that eq. \ref{vingt}
corresponds to a steady production of Minkowski quanta
during
the whole interacting period $\Delta \tau =T$ (infinite in eq. \ref{vingt})
we evaluate how many are produced.
(Contrary to the energy, the total number of Minkowski quanta
is a scalar under the Lorentz group)
 \begin{eqnarray}
\tilde N( \Delta \tau )  &=&  \int_{0}^{+\infty}  \! d \omega
\expect{\tilde 0_M}{ a_{ \omega} ^\dagger a_{ \omega}}{ \tilde 0_M}
\nonumber\\ &=&  \int_{0}^{+\infty}  \! d \omega
\expect{0_M}{\tilde  a_{ \omega} ^\dagger
 \tilde a_{ \omega}}{0_M}
\nonumber\\ &=&  \int_{0}^{+\infty}  \! d \omega
 \int_{-\infty}^{+\infty}  d\la\ |  \gamma_{\la,\om}( \Delta \tau )|^2
| \tilde \beta_{\la}|^2
\label{vun}
\end{eqnarray}
where $ |\tilde0_M>$ is the scattered (Schr\"odinger) state\footnote{
The simplest way to obtain this state is to find the scattering operator
$U$ such that $\tilde a_{\la,M}= U ^\dagger a_{\la,M} U$. Then $
|\tilde 0_M > =U |0_M >$.}.
The $ \tilde a_{ \omega}$ are related to the $ \tilde a_{\la,M}$
by (see eq. \ref{threethreevii})
\begin{equation}
 \tilde a_{ \omega} = \int_0^{\infty}\! d\omega\ \gamma_{\la,\om}
( \Delta \tau )  \tilde a_{\la,M}
\label{vdeux}
\end{equation}
where $\gamma_{\la,\om}( \Delta \tau )$ takes into account the time
dependence of the
coupling. As shown in \cite{pbt} $\gamma_{\la,\om}( \Delta \tau )$ is
non vanishing
only for the $\om$ which enter into resonance with the oscillator
frequency $m$ during the interaction period
 $ \tau_i <  \tau <
\tau_f = \tau_i + T$. When these frequencies belong to
\begin{equation}
 \omega_i = m e^{-a \tau_i} <  \omega <  m e^{-a \tau_f}= \omega_f
\label{vtrois}
\end{equation}
$ \gamma_{\la,\om}( \Delta \tau )$ may be
replaced by $ \gamma_{\la,\om}$ (given in eq. \ref{threethreeviii}).
Hence $ \tilde N( \Delta \tau )$ reads
\begin{eqnarray}
\tilde N( \Delta \tau ) &=&  \int_{ \omega_i}^{ \omega_f}  \!
{d \omega \over 2 \pi a \omega}  \int_{-\infty}^{+\infty}  d\la\ | \tilde
\beta_{\la}|^2
\nonumber\\ &=&  { \Delta \tau \over 2 \pi} \int_{-\infty}^{+\infty}  d\la\ |
\tilde
\beta_{\la}|^2
\label{vquatre}
\end{eqnarray}
The total energy emitted obtained from eq. \ref{vquatre} is
\begin{eqnarray}
\tilde H_M( \Delta \tau)  &=&  \int_{ \omega_i}^{ \omega_f}  \! { d
\omega \over 2 \pi a} \int_{-\infty}^{+\infty}  d\la\ | \tilde
\beta_{\la}|^2
\nonumber\\
&=& \int_{ \tau_i}^{ \tau_f}  \! { d \tau \over 2 \pi}
e^{-a\tau}
m \int_{-\infty}^{+\infty} \! d\la\ | \tilde
\beta_{\la}|^2
\nonumber\\ &=& \int_{V_i}^{V_f}  \! {d V \over 2 \pi}  {1 \over a^2V^2} m
\int_{-\infty}^{+\infty} \! d\la\ |
\tilde
\beta_{\la}|^2
\label{vcinq}
\end{eqnarray}
in perfect agreement with eq. \ref{vingt} if the frequency width
of the oscillator in small compared to $ m$.
The rate of production (eq. \ref{vquatre} divided by $ \Delta \tau$) is
(small width limit) ${e^2} \alpha _{m}^2
\beta_{m}^2$ which is the rate of jumps for an inertial
oscillator in a bath at temperature $a/2 \pi$. Therefore the number of
Minkowski quanta
produced by the thermalised oscillator equals the number of internal jumps.

We now generalise these results to an arbitrary linear coupling.
We
believe
that it can be generalised, using the same type of argumentation,
 to nonlinear couplings as well. The
proof goes as follow. Any scattering of Rindler quanta by an accelerated
system which leads to a thermal equilibrium during a time much larger
than
$1/a$ can be described as in eq. \ref{onze} by
\begin{equation}
\tilde a_{\la,R} =  S_{\la \la^\prime} a_{\la^\prime,R}
 \label{vsix}
\end{equation}
where repeated indices are summed (or integrated) over and where the
summation over $\la^\prime$ includes both $u$ and $v$-modes (as in
eq. \ref{neuf}).
 The matrix $S$ satisfy the
unitary relation
\begin{equation}
 S_{\la \la^{\prime\prime}}  S_{\la^{\prime\prime} \la^\prime}^\dagger
 = \delta _{\la \la^\prime}
 \label{vseven}
\end{equation}
which express the conservation of the number of Rindler quanta since
$S_{\la, \la^\prime}$ mixes positive Rindler frequencies only.
It is convenient to introduce the matrix $T$ (from now on we
do not write the indices)
\begin{equation}
S=1+iT
 \label{veight}
\end{equation}
which satisfies
\begin{equation}
2 Im T = TT^\dagger
\label{vnine}
\end{equation}
We introduce also the vector operator $b= \left( a_{\la,R};a_{\la,L};
a_{\la,R}^\dagger;a_{\la,L}^\dagger \right) $. Then eq. \ref{vsix}
 can be written as
\begin{equation}
\tilde b= {\cal  S} b
\label{trente}
\end{equation}
where $ {\cal S}$ has the following block structure
\begin{equation}
 {\cal  S} =
\left( \begin{array}{cccc}
1+iT & 0 & 0 & 0 \\
0 & 1 & 0 & 0 \\
0 & 0 & 1-iT^\dagger & 0 \\
0 & 0 & 0 & 1
\end{array} \right)
\label{tone}
\end{equation}
since the $u$ and $v$-modes on the left quadrant are still free.
On the other hand, the Bogoljubov transformation eq. \ref{threethreex}
 reads in this
notation
\begin{equation}
c ={ \cal  B} b
\label{ttwo}
\end{equation}
where $c=  \left( a_{\la,M};a_{- \la,M};a_{\la,M}^\dagger;a_{- \la,M}^\dagger
\right)$
 and where $ {\cal  B}$ is
\begin{equation}
  {\cal  B} =
\left( \begin{array}{cccc}
 \alpha &  0 & 0 & - \beta \\
 0 & \alpha & - \beta  & 0 \\
 0 & - \beta &\alpha & 0 \\
 - \beta  & 0 & 0 & \alpha \\
\end{array} \right)
\label{tthree}
\end{equation}
the diagonal matrices (in $\la$) $\alpha $ and $\beta$ being taken real.
The scattered Minkowski operators are given by
\begin{equation}
\tilde c =  {\cal  B}{ \cal  S}{ \cal  B }^{-1} c = \left(
{ \cal  S} + {\cal  B}\left[ { \cal
S},{ \cal  B }^{-1} \right]_- \right) c = {\cal  S}_M c
\label{tfour}
\end{equation}
Since  ${\cal  S}$ and  ${\cal  B}$ do not commute,  ${\cal  S}_M$ has non
diagonal elements which encode the production: \begin{equation}
  {\cal  S}_M  =
\left( \begin{array}{cccc}
 \tilde \alpha_1 &  0 & 0 & - \tilde \beta _1 \\
 0 & \tilde  \alpha_2 & \tilde  \beta^\dagger _1  & 0 \\
 0 &  \beta^\dagger_2 &\alpha _1^\dagger & 0 \\
 - \tilde  \beta_2  & 0 & 0 &\tilde  \alpha^\dagger_2 \\
\end{array} \right)
\label{tfive}
\end{equation}

where the $\tilde \alpha$ $\tilde \beta$ are given in terms of $T$ by
(see eq. \ref{dhuit})
\begin{eqnarray}
 \tilde  \alpha _1 &=&  1+i  \alpha T  \alpha
\nonumber\\
 \tilde
\beta _1  &=& -i  \alpha T \beta
\nonumber\\
\tilde  \alpha _2 &=& 1 +i \beta T^\dagger  \beta
\nonumber\\
\tilde
\beta_2 &=& i  \beta  T  \alpha
\label{tsix}
\end{eqnarray}
QED

\section{The Black Hole}
\subsection{The kinematics of the collapse and the scattered modes}

We shall work in the background metric of a spherically symmetric
collapsing
star of mass $M$. Outside the star the geometry is described  by the
Schwarzshild metric
\begin{eqnarray}
ds^2 &=& (1- {2 M \over r}) dt^2 - (1- {2 M
\over r})^{-1} dr^2 - r^2 d \Omega^2 \nonumber\\
&=& (1- {2 M \over r}) du dv -r^2 d \Omega^2 \nonumber\\
v,u &=& t \pm r^*\nonumber\\
r^* &=& r + 2M \ln{ r-2M \over 2M}
\label{bh1}
\end{eqnarray}

The specific  collapse we consider is
produced by a spherically symmetric shell of pressureless
massless matter. Inside the shell space is
flat and the metric reads
\begin{eqnarray}
ds^2 &=& d \tau ^2 - dr^2 - r^2 d \Omega^2 \nonumber\\
&=& dU dv -r^2 d \Omega^2 \nonumber\\
v,U &=& \tau \pm r
\label{bh3}
\end{eqnarray}
where $v$ is the same coordinate in eq. \ref{bh1} and eq. \ref{bh3}
since on ${\cal I}^-$ space time is flat on both sides.
The collapsing shell, taken to be thin, follows the geodesic $v=v_S$.
The connection between the two metrics is obtained by imposing the
continuity of $r$ along the shell's trajectory
\begin{equation}
dU = du (1 - {2 M \over r(u,v_S)}) = du (1 - {4 M \over v_S - U})
\end{equation}
Then by choosing $v_S=4M$ one gets
\begin{eqnarray}
du &=& - { dU \over U} \left( 4M - U
\right)
\nonumber\\  u(U) &=& U - 4M \ln ({- U \over 4M})
\end{eqnarray}

In the static space time outside the star, the Klein-Gordon equation
for a mode of the form $ \varphi_{l,m}
= {1 \over \sqrt {4 \pi r^2}} Y_{lm}(\theta, \varphi)
\psi_{l}(t,r)
$
reads
\begin{equation}
\left[ \partial_t^2 -\partial_{r^*}^2 - (1- {2 M \over
r}) \left[ {l(l+1) \over r^2} + m^2 + { 2 M \over r^3} \right] \right]
\psi_l
(t,r) =0
\label{bh2}
\end{equation}
Near the horizon $r-2M << 2M$, it  becomes the wave equation
for a massless field
in $1+1$
dimensions.
By considering
only
the s-wave sector of
a massless field
and dropping the
residual
 "quantum potential" ${2M(r-2M) / r^4}$ the
conformal invariance holds everywhere, inside as well as outside
the
star. (This does not depend on our specific collapse: it is also valid if
one assumes, following Hawking{ \cite{hawk2}}
, that the geometrical optics limit is valid
inside the star.)
 From now  on we shall work in this simplified
context and only discuss
briefly the differences with the more realistic four dimensional case.

The Heisenberg state is chosen to be the initial vacuum
i.e. vacuum with respect to the modes which have positive
$v$-frequency on ${\cal I}^-$. Those modes are reflected at $r=0$ and read
\begin{equation}
\varphi_{\omega,0,0}(v,u) = {1 \over 4 \pi r \sqrt{\omega}}
\left( e^{-i \omega v} - e^{-i\omega U(u)} \right)
\end{equation}
Hence, for $u > 4M$ (or $-M < U < M$, on both sides of the horizon)
 the state of the field tends exponentially quickly (in $u$) to Unruh
vacuum, i.e.
vacuum with respect to the modes
\begin{equation}
\exp(-i \omega v) \quad {\hbox{and}}\quad \exp({i\omega \over 4
M} e^{-u\over 4
M})
\end{equation}

The Schwarzschild $u$-modes  $\chi_{\la}(u)= {e^{-i\la u}}/( 4 \pi r
\sqrt{
\la}) $
are needed to analyse
the particle content
of the scattered modes
$\varphi_{\omega}$
on ${\cal I^+}$. In terms of $U$ they take the form
\begin{equation}
\chi_{\la}(u) =  \theta(- U){1 \over  4 \pi r
\sqrt{\la}}
 ({-U \over 4M} )^{i \la 4 M} e^{- i \la U}
\end{equation}
The exact Bogoljubov coefficients between $\varphi_\omega$ and $\chi_{\la}$
are given by
\begin{equation}
\alpha_{\omega,\la} =
\bk{\varphi_\omega,\chi_{\la}}
= {1 \over 4 \pi}
\sqrt{ \omega \over \la}
\Gamma ( 1 + i 4 M \la)
[ 4 M ( \omega - \la)]^{-i 4 M \la}
e^{\pm 2 \pi M \la}
\end{equation}
where the $\pm$ is to be understood as $+$ if $ \omega > \la$ and $-$
if $ \omega < \la$. The expression for $\beta_{\omega,\la}$ is obtained by
taking $\la$ into $-\la$. The asymptotic Bogoljubov coefficients
(relating Kruskal modes to Schwarzschild modes
eq. \ref{threethreevii} et seq. )
are recovered in the
limit $\omega \to + \infty$ since it corresponds to resonance at late times
$u \to +\infty$ (see eq. \ref{vtrois}). In this limit
the black hole emits quanta at the Hawking temperature $1/8 \pi M$
 since $\vert \beta_{\omega,\la}/ \alpha_{\omega,\la}\vert ^2 =
e^{-8 \pi M \la} $.

Having described the kinematics of the collapse we now turn to the
post-selection of the emitted quanta.
The
new difficulty lies in the
renormalisation of the energy momentum tensor which must be carried out
in curved space times. We therefore turn to this point.

\subsection{Weak-values in curved space-time}

Wald has proposed a set of eminently reasonable
conditions that a renormalised energy
momentum operator should satisfy \cite{wald}.
By an argument  similar to
Wald's (or
simply by verifying that it is in accord with his axioms), it is
 possible to deduce that  $T_{\mu \nu ({\rm ren})} (x)$ can
be
written in the following way
\begin{equation}
T_{\mu \nu ({\rm ren})} (x) =
T_{\mu \nu} (x) - t_{\mu \nu ({\rm S})} (x) I
\end{equation}
where $T_{\mu
\nu} (x)$ is the bare energy momentum tensor.  The subtraction term  $t_{\mu
\nu ({\rm S})} (x)$ is an (infinite) conserved c-number function only of the
geometry at $x$. It can be understood \cite{mpblocal} \cite{mas}
as the (infinite) ground state
energy of the "local inertial vacuum": that state which most resembles
Minkowski vacuum at $x$. Numerous techniques have been
developed
to
calculate $t_{\mu \nu ({\rm S})}$ and we refer the reader to
\cite{birreld}
for a
review.

In a state, say the Heisenberg vacuum
$\ket{0}$, the expectation value of $T_{\mu\nu}$
 takes the form
\begin{equation}
 \expect{0}{T_{\mu \nu
({\rm ren})} (x)}{0} = \expect{0}{T_{\mu \nu } (x)}{0} -  t_{\mu \nu
({\rm S})} (x)
\label{beware}
\end{equation}
 where both terms on the r.h.s. are infinite but
their difference is finite.

In a pre- and post-selected ensemble, the weak values of $T_{\mu \nu}$
reads
\begin{equation}
T_{\mu \nu ({\rm weak})} = {
\expect{0}{\Pi T_{\mu \nu ({\rm ren})} } {0} \over  \expect{0}{\Pi }
{0} }
 \label{bh11}
\end{equation}
where $\Pi$ is the projector (or more generally the self
adjoint operator) that realises the post-selection. Inserting eq. \ref{bh11}
 into this
expression yields
 \begin{equation}
T_{\mu \nu ({\rm weak})}(x) = {
\expect{0}{\Pi {T_{\mu \nu }}(x) } {0} \over  \expect{0}{\Pi } {0} } -
t_{\mu \nu ({\rm S})}(x)
\label{bh12}
\end{equation}
By expressing ${T_{\mu \nu }}(x)$ in terms of the operators which
annihilate the Heisenberg vacuum one obtains
\begin{equation}
T_{\mu \nu ({\rm weak})}(x) = \int_0^\infty\!\! d \omega
\int_0^\infty\!\! d \omega^{\prime} {
\expect{0}{\Pi a_{\omega}^{\dagger} a_{\omega^{\prime}}^{\dagger}
 } {0} \over  \expect{0}{\Pi } {0} } \hat T_{\mu \nu }
(x)\left[ \varphi^*_\omega \varphi^*_{\omega^{\prime}}
\right] +
\expect{0}{T_{\mu \nu
({\rm ren})} (x)}{0}
\label{2terms}
\end{equation}
where $ \hat T_{\mu \nu } (x)$
is the classical differential operator
which acting on the waves $ \varphi^*_\omega$ gives their energy
density. The first term depends on the particle content of the
post-selection and the second one is the energy density of the
Heisenberg vacuum
eq. \ref{beware}.

The formula eq. \ref{bh12}
 warrants a few additional comments. First notice that their are
parts of $\bk{T_{\mu\nu}}_{weak}$
that are
entirely contained in the subtraction.
Most notably there is the trace anomaly and those components of the
energy
momentum tensor which are related to it by energy conservation
(in two dimensions they are $T_{uu,v}$ and $T_{vv,u}$). These parts
are
independent of the post-selection or, expressed differently, do not
fluctuate.

An additional (and related)  feature which has already been mentioned in
chapter 3 concerns the absence of correlations between $T_{uu}$ and $T_{vv}$.
Not only shall this give rise to the particular structure of vacuum
fluctuations that extend back to ${\cal I}^-$, but it also implies that on the
horizon the in-going flow and the out-going flow fluctuate independently (for
instance the post selection of an outgoing particle on ${\cal I}^+$ does not
affect $T_{vv}$ outside the star and in particular
 on the horizon $r=2M$). This last effect disappears partially
when considering the potential barrier that occurs in the wave equation
eq. \ref{bh2}.

\subsection{The different post-selections}

Since an external observer does not have access to the region of space time
beyond  the horizon, the post-selections that he can perform are
restricted to an incomplete ($U<0$) region of space time and are therefore
incomplete as well.

The post-selection could consist in specifying the state of the
outgoing photons outside the star. For instance  one could specify that the
state $\Pi \ket{0}$
be $a^\dagger_\la \ket{B}$ where $\ket{B}$ is Boulware vacuum and
$a^\dagger_\la$ creates a Schwarzshild photon.
The corresponding weak value
reads
\begin{eqnarray}
&\ & T_{\mu \nu ({\rm weak})}(x) =
{2 \over \alpha_\la \beta_\la } \hat T_{\mu \nu }(x)
\left[ \varphi^*_{\la,K} \varphi^*_{-\la,K} \right]
\nonumber
\\
 &\ &\quad + \left[ {
\expect{B} {T_{\mu \nu } (x)} {0} \over  \scal{B}{0}}
- \expect{0}{T_{\mu \nu} (x)}{0} \right] +
\expect{0}{T_{\mu \nu
({\rm ren})} (x)}{0}
\label{bh18}
\end{eqnarray}
where the new Kruskal modes $\varphi_{\la,K}$ are defined as in
eq. \ref{threethreevii}.
The first term is equal to the energy of the photon $\la$, the second
one is
the difference of energy between Boulware vacuum and the
Heisenberg vacuum
and the third one
is the Heisenberg vacuum energy eq. \ref{beware}. The second term appears
because one has specified that, apart from $\la$, their is no other photon
emitted. This is why this term is singular on the horizon.
In addition, the probability to obtain the state $a^\dagger_\la
\ket{B}$ vanishes in the absence of backreaction and is
in the semiclassical approximation of  order
 $e^{-M^2}$ where $M^2$ is approximately
the total number of photons emitted.

An alternative post-selection consists in tracing over all the photons
except the photon $\la$ which is imposed to be present (in the Rindler
problem this corresponds to the projector eq. \ref{threefourvii}).
 Then the weak value is
simply
\begin{equation}
T_{\mu \nu ({\rm weak})}(x) = {2 \over \alpha_\la \beta_\la } \hat T_{\mu
\nu } (x)
\left[ \varphi^*_{\la,K} \varphi^*_{-\la,K} \right]+ \expect{0}{T_{\mu
\nu
({\rm ren})} (x)}{0}
\label{bh19}
\end{equation}
We could  post-select a wave packet rather than a mode of fixed energy $\la$ in
which case eq. \ref{bh19} would be finite on the horizon.

Having traced over all
the other photons, the second term of eq. \ref{bh18} is absent in
eq. \ref{bh19}. Nevertheless it can also be constructed as the sum of weak
values that
specify completely the state times the probability that they occur (in
similar manner to the unitarity relation eq. \ref{weakix}).
 In this way the difference of
energy between Boulware vacuum and the Heisenberg vacuum is realised as the
sum over all possible radiated photons times the thermal probabilities that
they occur.

Finally we consider post-selection by an inertial
two level atom at large distance
from the black hole. In this case  the final state is  partially
specified, since the detector is coupled to a finite set of modes, and one
 therefore obtains a result
similar to eq. \ref{bh19} wherein the first term consists in the
post-selected Hawking photon. We shall display its properties
in the next section.

It is
also
interesting to speculate
about the
nature of the in-going vacuum fluctuations. These could be analysed by
post-selecting the presence of ingoing quanta near the horizon.
A ''natural" set of
 modes to post-select near the horizon are Kruskal $v$-modes. One is
therefore led to consider the Kruskal vacuum fluctuations in Schwarzshild
vacuum, which is similar to considering Minkowski fluctuations in Rindler
vacuum. If space time were the full Schwarzshild manifold, these would
present a singularity on the past horizon that could be smoothed out using
wave packets. Since space time is not the full
Schwarzshild manifold (there is no past horizon)
the star's surface
will play the role of past
horizon and one expects large energy densities
in the outermost layers of the star.

\subsection{From vacuum fluctuations to black hole radiation}

We now turn to that piece of the weak value
which depends on the particular mode post-selected by the two level atom.
This piece in completely independent of the geometry for the $s$-wave
that we are considering since we neglect the residual potential of the
dalembertian  eq. \ref{bh2}. Hence
the mapping of the results obtained in the Rindler problem to the
black hole is straightforward.
Let us choose the time dependant coupling $f(t)$ of our detector
such that it will be excited only by spherical photons centred around
$u=u_0$ with (Schwarzshild) energy $\la = m$. Such an example
of wave packet is offered by the Fourier
components
given in eq. \ref{threefivexxv}
\begin{equation}
 c_\la = D { \la \over m} e^{i \la u_0}e^{-(\la - m )^2 T^2 /2}
 (1 -
e^{-2\pi \la /a})
\end{equation}
The spread in time is $\Delta t = \Delta u = T$
 and $u_0$ is taken well inside
the region $u>0$
 where the isomorphism of the scattered waves and the
Kruskal modes is achieved.

The picture that emerges, if the two level atom is found excited after
the switch off, is that
this photon
results form a spherically symmetric vacuum fluctuation on ${\cal
I}^-$ which carries zero total energy and is located in a region
\begin{equation}
\vert v  -
v_\infty\vert=\vert \Delta U\vert=\vert {\Delta u e^{- u_0 / 4 M}}\vert
\simeq T e^{- u_0 / 4 M}
\label{blaack}
\end{equation}
where $v = v_\infty$ ($=0$ in our
collapse) is the light ray that
shall become the
future horizon $U=0$.
Indeed this localisation is furnished by the $v$ dependence of the
weak value
 on ${\cal I}^-$ which reads (see equation eq. \ref{weekiv})
\begin{equation}
\bk{T_{vv}(\cal I^-)}_{\rm weak}
\simeq  {1 \over 4 \pi r^2}{16 M^2 \over v^2} {m  \over  2 \sqrt{\pi} T }
(N_{m} +1)\ \exp \left[
-\left[
{ 4 M  \over T}  (  \ln ({-v-i\e
\over 4M}) + u_0 )
\right]^2\right]
\end{equation}
This results from the fact that the analysis by an inertial observer near
${\cal I}^-$ is isomorphic with what was called the Minkowski interpretation
in chapter 4.
 As in the accelerated case, the energy density is
enhanced
by the jacobian $du/dU= e^{u / 4 M}$ centered around $u=u_0$ which appears here
as
$1/v^2$ when the reflection at $r=0$ is taken into account. Hence after a
$u$-time of the order of $4M \ln M$, the energy density in $T_{vv}$
(rescaled by $ 4 \pi r^2$)
become "transplanckian" and located within a "cisplanckian"
distance $\Delta v$ (If one does not rescale $T_{\mu\nu}$
the transplanckian energies only exist in a region of finite $r$).
The analysis and the consequences of these transplanckian energies is
presented in a separate paper \cite{EMP}. In that article it is argued that
the nonlinearity of general relativity cannot accommodate these densities and
that  a taming mechanism must exist if
Hawking radiation does exist.

After issuing from ${\cal I}^-$, the vacuum fluctuation
 contracts until it reaches $r=0$ and then reexpands along $U=const$
lines. Upon crossing the
surface of the star in a region $\Delta U$ centred on the horizon,
it separates
into a piece (the partner)
that
falls into the singularity,
carrying a negative Schwarzshild
energy equal to $- m$,
and a piece carrying positive
energy equal to $ m$ that keeps expanding and escapes to ${\cal
I}^+$ to constitute the post-selected quantum that will induce the
transition (see eq. \ref{weeki} and figure 1).
The analysis performed by an inertial asymptotic observer near the
detector, on ${\cal
I}^+$,
is isomorphic with
the Rindler interpretation of chapter 4.

If the two level atom is found
in its ground state
 after the switch off,
its wave function
is correlated to the absence of the Hawking photon specified by $c_\la$.
In that case, one would find near ${\cal I}^-$ a vacuum fluctuation whose
energy content is exactly the opposite of the previously considered
case (times $N_{m}$).
Near ${\cal I}^+$ it would contain a negative energy flux of total energy $-
N_{m} m$ encoding the fact that their is one quanta absent from the
thermal flux emitted by the black hole.

If more realistically, we take
a two-level atom coupled locally to the field
(i.e. coupled to all the modes $l>0$),
 it
will
 post select
particles coming out of the black hole in its direction. Then the picture
that emerges is essentially the same as for
an s-wave
except that on ${\cal I}^-$ the vacuum fluctuation is localised
on the antipodal point of the detector.
The  created quantum and its partner,
 are on the same side and
not antipodal
(with respect to each other)
because they have opposite energy.

We now turn to the description in the intermediate regions in order to
interpolate between the descriptions between
${\cal I}^-$ and ${\cal I}^+$.
One possible interpolation
consists in using a set of static observers at constant $r$.
Then the ''Rindler" description would be used everywhere outside the
star. However a difficulty arrises in this scheme if one really considers a
set of material ''fiducial" \cite{sus1}
detectors
at constant $r$. For upon interacting
with the field and thermalising at the local temperature
$\sqrt{r\over(r-2M)}{1 \over 8 \pi M}$ the
detectors will emit large amounts of
ultraviolet Kruskal
"real"
quanta (see chapter 5 wherein it is shown how the accelerated atom
transforms vacuum fluctuations into "real" quanta). The backreaction of
these
quanta
cannot be neglected and, as already stated, cannot be evaluated
owing to the transplanckian energy
they carry.

An alternative interpolation
 consists in giving the value of $T_{\mu\nu}$ in the local
inertial coordinate system (Riemann normal coordinates). This stems from the
idea that
local
physics
should be describe locally in such a coordinate system. This approach has
 been used
in defining the subtraction necessary to
renormalize the energy momentum tensor \cite{mpblocal} \cite{mas}.
In the two dimensional model the local inertial coordinates are easy to
construct. Since $\tilde u = r(u,v)$ is an affine parameter along the
geodesics
$v={\rm constant}$, a natural way to represent the outgoing flux outside the
star is as
\begin{equation}
T_{\tilde u \tilde u} (\tilde u) = \left( {d u(r,v) \over dr} \right )^2
T_{uu}(u(r,v))
\end{equation}
This is represented in both a
Penrose diagram and Eddington-Finkelstein coordinates in figures 4 and 5.

After the Hawking photon reaches a distance $ r \geq 4M$ it travels in
flat space,
it
is no longer modified
and the backreaction may be safely computed from
 $\bk{T_{\mu\nu}}_{weak}$. But a $v$ time of order $4M \ln M$
before it reached flat space the photon already carried planckian energy
densities in this local description.

To obtain a first indication of the gravitational backreaction we consider
the linear modification of the metric $\delta g_{\mu\nu}$ and describe it
quantum mechanically. In first order perturbation theory
$\delta g_{\mu\nu}$ itself could be taken to be the additional system (the
weak detector) introduced in chapter 2. Then the weak
value of $\delta g_{\mu\nu}$ is obtained by integrating Einstein's equations
with $\bk{T_{\mu\nu}}_{weak}$ as source since the weak values obey
the Heisenberg equations of motion.
For s-waves the constrained part of the metric only will be modified.
Futhermore since the total energy carried by the weak value of $T_{\mu\nu}$
vanishes from ${\cal I}^-$ till the emergence of the fluctuation from the star
after reflection on $r=0$, the weak value of $\delta g_{\mu\nu}$ will vanish
outside the interval $\Delta v$ eq. \ref{blaack}.
Within that interval the precise
shape of  $\delta g_{\mu\nu\ weak}$ will depend on the particular choice of
post-selected wave packet. On the contrary, outside the star, for $r>4M$ and
$u> u_0$, the weak value of $\delta g_{\mu\nu}$ will encode the mass loss
$\omega$ and in fact describes a new Schwarzshild space where the mass is
$M-\omega$.

One can also consider the backreaction of the Hawking photon onto itself and
onto the subsequent photons. This self interaction is governed by a
hamiltonian of the form $H_{int} = T_{\mu \nu} D^{\mu \nu
\alpha\beta}T_{\alpha\beta}$ where $D$ is the linearized gravitation
propagator. In this approximation the backreaction is given by
$\bk {H_{int}}_{weak}$.
But since $\bk{T_{\mu\nu}}_{weak}$ becomes larger than $1$ (in Planck units)
when the selected photon is a Planck distance from the horizon the linear
approximation invariably fails. This is compounded by the non
renormalisability of the gravitional interaction which presumably does not
lead to an asymptoticly free theory beyond the Planck scale. How Hawking
radiation could still be realised in a consistent theory of gravity remains
to be seen.
\\
\\
{\bf Acknowledgements.}
\noindent The authors would like to thank R. Brout,
F. Englert, S. Popescu and Ph. Spindel for very helpful
discussions.

\vfill \newpage

Figure Captions

Figure 1.

\noindent The weak value of $T_{vv}$ if the two level
atom gets excited is represented for the values of
parameters $m=2a$ and $T = 3 a^{-1}$. The $v$ axis is
given in units of $a^{-1}$. On top of the figure
the absolute value of the coupling function $f(\tau)$ of
the two level atom to the field is represented. Underneath
is a graph of  $\bk{T_{vv}(I_+, V < 0 )}_{weak\  e}$ ($=
\bk{T_{vv}(I_-, V < 0 )}_{weak\  e}$ by causality) and
$\bk{T_{vv}(I_+, V > 0 )}_{weak\  e}$. Notice that the scale
of this last drawing is defferent from the others since
this weak value is proportional to $N_m$ whereas the
others are proportional to $1 + N_m$. Underneath
 the real and
imaginary part of
$\bk{T_{vv}(I_-, V > 0 )}_{weak\  e}$ are represented. These
pictures show how a Rindler observer would see the
weak values.

Figure 2.

\noindent
The real and imaginary parts of
$\bk{T_{VV}(I_-)}_{weak\  e}$ is represented for the same
values of parameters as in figure 1. The $V$ axis is
given in units of $a^{-1}$. $T_{VV}$ presents
very strong oscillations near $V=0$ which are not represented.
If one
considers only
${\rm Re} [\bk{T_{VV}(I_-)}_{weak\  e}]$ for $V>0$ and compares
it to ${\rm Re} [\bk{T_{vv}(I_-, V>0)}_{weak\  e}]$ of figure
1, then the positive hump to the left of $V=1$
corresponds to the central positive hump centered on
$v=0$ and the negative oscillations to the left of
$V=0.2$ correspond to the dip between $-7 < v <-2$. The tail
oscillations in the Rindler description are enhanced by
the jacobian that passes from Rindler to Minkowski
coordinates in such a way that the integral of
the graphs in figure 2 vanish.

Figure 3.

\noindent
The mean energy emitted to order $g^2$ at thermal
equilibrium (rescaled by the probility to
emit a left photon) is represented for
$m=2a$ and $T=3a^{-1}$ from
both the Minkowski and Rindler point of vue and for
$m=2a$ , $T=10a^{-1}$ from  the Rindler point of vue only.
It is apparent that as $T$ increases the Rindler energy
emitted per transition of the atom tends to zero. In the
Minkowski description the tails of postive Rindler energy
are enhanced by the jacobian $dV/dv$ to make the total
Minkowski energy emitted postive.

Figure 4.

\noindent The local description of a vacuum fluctuation
giving rise to a Hawking photon emitted around $u=u_0$ is
represented in a Penrose diagram. The shaded areas
correspond to the regions where
$T_{\tilde u \tilde u}(\tilde u)$ is non vanishing.
$v=v_S$ is the trajectory of the collapsing spherically
symmetric shell of massless matter.

Figure 5.

\noindent The same as in figure 4 drawn in advanced
Eddington-Finkelstein coordinates $(r,v=t+r^*)$.

\end{document}